\documentclass[5p,number,sort&compress]{elsarticle}

\usepackage{units}
\usepackage{amsmath}
\usepackage{amssymb}
\usepackage{wasysym}

\newcommand{\hess}{H.E.S.S.}  
\newcommand{\hgc}{HESS~J1745-290}
\newcommand{\astar}{Sgr~A$^*$} 
\newcommand{\aeast}{Sgr~A~East}

\newcommand{\gr}{$\gamma$-ray}
\newcommand{\grs}{$\gamma$-rays}

\journal{Astroparticle Physics}

\begin{document}

\begin{frontmatter}

\title{Gamma rays from the Galactic Centre Region: a review}

\author{Christopher van Eldik}
\address{ECAP, University of Erlangen-N\"urnberg, Erwin-Rommel-Str. 1, D-91058 Erlangen, Germany}
\ead{Christopher.van.Eldik@physik.uni-erlangen.de}

\begin{abstract}
During the last decades, increasingly precise astronomical observations of the Galactic Centre (GC) region at radio, infrared, and X-ray wavelengths laid the foundations to a detailed understanding of the high energy astroparticle physics of this most remarkable location in the Galaxy. Recently, observations of this region in high energy (HE, $\unit[10]{MeV} - \unit[100]{GeV}$) and very high energy (VHE, $>\unit[100]{GeV}$) \grs\ added important insights to the emerging picture of the Galactic nucleus as a most violent and active region where acceleration of particles to very high energies -- possibly up to a \unit{PeV} -- and their transport can be studied in great detail. Moreover, the inner Galaxy is believed to host large concentrations of dark matter (DM), and is therefore one of the prime targets for the indirect search for \grs\ from annihilating or decaying dark matter particles. In this article, the current understanding of the \gr\ emission emanating from the GC is summarised and the results of recent DM searches in HE and VHE \grs\ are reviewed.
\end{abstract}


\end{frontmatter}

\tableofcontents

\section{Introduction}
Since the discovery of the Galactic nucleus by radio observations in the 1950s \cite{Piddington:1951, McGee:1954}, the GC has been subject to intense astrophysics research. Because of its proximity to Earth, it serves as a unique laboratory for investigating the astrophysics of galactic nuclei in general. On the one hand, the dynamics of inner (roughly \unit[300]{pc} in radius) region of the Galaxy is largely driven by the presence of the supermassive black hole, identified with the strong radio source \astar\ \cite{Balick:1974}. On the other hand, there is a wealth of other interesting phenomena concentrated in this region, such as active star forming regions emitting powerful winds, supernova remnants (SNRs, partly in contact with molecular material), pulsar wind nebulae (PWNe), populations of high-mass X-ray binaries and large-scale non-thermal filaments (for a review, see e.g.\ \cite{Mezger:1996, Morris:1996}). In addition, the interstellar medium (ISM) is on average about an order of magnitude denser than in other parts of the Galactic disk, and the region is pervaded by strong magnetic fields, probably by far exceeding the level of $\unit[50]{\mu G}$ \cite{Crocker:2010} (as compared to typically a few $\unit{\mu G}$ in the disk \cite{Beck:2013}).

Obscured by a thick layer of dust along the Galactic plane, the inner part of the Milky Way, including the Galactic nucleus with its multi-million solar mass black hole (BH), is escaping observations with classical optical telescopes. It is, however, visible in a broad range of other frequencies of the electromagnetic spectrum, since the dust torus is mostly transparent to low energy (radio and far infrared) radiation as well as to high-energy photons (X-rays up to multi-TeV \grs). While far infrared (IR) emission is a tracer of sites of ongoing star formation (the strong optical and UV emission from massive, young stars is converted by dust into IR radiation \cite{Devereux:1990}), non-thermal radio, X-ray and \gr\ emission traces populations of charged particles that underwent acceleration to supra-thermal energies in cosmic accelerators such as SNRs, PWNe, or in colliding winds driven by massive stars in star forming regions (see, e.g., \cite{Hinton:2009}). Non-thermal radio and X-ray emission is produced by multi-GeV to multi-TeV electrons, emitting synchrotron radiation due to interaction with surrounding magnetic fields, while HE and VHE \grs\ result from inverse Compton (IC) scattering or bremsstrahlung of electrons off ambient photons or gas, respectively, or from inelastic collisions of high energy cosmic rays (mostly protons) with the ambient medium, and successive production of \grs\ via decay of neutral pions \cite{longair:2011}. It is therefore clear that imaging the GC region at these photon energies yields a completely different view than the one that would be seen by an optical telescope.

During the last decade, a new generation of HE and VHE $\gamma$-ray telescopes enabled a very detailed study of the various high-energy processes taking place in the GC region. The Large Area Telescope (LAT, \cite{LAT:2009}, onboard the Fermi satellite) and the currently operated ground-based telescope arrays H.E.S.S. \cite{Hinton:2004}, MAGIC \cite{Aleksic:2012} and Veritas \cite{Holder:2008} outperform 1st generation instruments like EGRET, Whipple and HEGRA by far, not only due to their larger collection area, but also in terms of angular resolution, energy resolution, and 
background suppression. The resulting boost in $\gamma$-ray detection sensitivity has increased the number of discovered HE and VHE $\gamma$-ray sources (as of mid 2014, 1873 HE \cite{Nolan:2011} and 150 VHE \cite{TeVCAT} $\gamma$-ray sources are known), and has also enabled detailed studies of the energy spectra, the spatial morphology and temporal variability of these sources. What concerns the GC region, several new sources of high-energy $\gamma$-ray emission have been discovered and characterised by these instruments. These sources are key to understand the high-energy astrophysical phenomena at work in that region on both small scales ($\leq \unit[10]{pc}$, the typical spatial resolution of current instruments) and large scales ($\geq \unit[300]{pc}$, exceeding even the size of the GC ridge).

Furthermore, GC observations at high energies serve the purpose of unveiling the nature of dark matter (DM). As the Milky Way is believed to be embedded in a halo of DM particles, and their density is expected to be strongly peaked towards the GC, the GC region is an excellent place to search for the annihilation of DM into standard model particles. Since typical DM masses are expected to lie in the range of \unit[10]{GeV} to several \unit[10]{TeV}, HE and VHE $\gamma$-ray instruments are in general well suited to detect $\gamma$-rays produced in the annihilation process. No firm discovery of DM annihilations has been made yet. However, observations of the inner Galactic halo provide limits on the DM annihilation cross section that are among the strongest obtained so far for this range of DM particles masses.

This review gives an overview, based on $\gamma$-ray observations, about the current knowledge on the high-energy astrophysics of the central few \unit[100]{pc} region of the Milky Way and summarises the progress made in indirect searches for DM annihilations using $\gamma$-rays. It is not meant to provide a complete picture on the astrophysics of the region, but discusses from an observational point of view key aspects addressed by recent HE and VHE $\gamma$-ray data. In section \ref{sec:stage}, we first provide an overview about the multi-wavelength picture of the region, before we review recent $\gamma$-ray observations in the context of high-energy astrophysics (section \ref{sec:part1}) and indirect DM searches (section \ref{sec:part2}). We end by presenting a short outlook towards observations of the GC with future observatories, concentrating on the proposed next generation $\gamma$-ray instruments.

\section{Setting the stage: a multi-wavelength view of the Galactic Centre region}
\label{sec:stage}

\subsection{Distance to the Galactic Centre}
Based on different measures (such as tracing the orbits of the S-star population orbiting \astar \cite{Gillessen:2009, Eisenhauer:2005cv}, using the spatial distribution of globular clusters in the Milky Way \cite{Bica:2006}, or taking advantage of the known absolute magnitudes of red clump stars \cite{Nishiyama:2006}, RR Lyrae stars and  cepheids \cite{Groenewegen:2008}), the distance to the GC is about $R_0\sim\unit[7.5-8.5]{kpc}$, although values of up to $\sim\unit[9]{kpc}$ have been claimed in earlier works. As the exact distance is only of minor importance for the physics results discussed in this work, a canonical value of $R_0=\unit[8]{kpc}$ \cite{Reid:1993fx} is used throughout this review. This means that an angular distance of $1^\circ$ on the sky translates into a projected linear distance of $\sim\unit[140]{pc}$ at the GC position. In passing, we note that given the typical angular resolution of current $\gamma$-ray instruments ($\sim 0.05^\circ - 0.2^\circ$, strongly dependent on the $\gamma$-ray energy), the smallest physical sizes that can be probed by $\gamma$-ray observations at the GC distance are therefore of the order of $\unit[10]{pc}$\footnote{Note that this hold for steady emission. For rapidly variable emission, the smallest sizes $\Delta s = c\Delta t$ (characterised by the maximum distance between causally connected regions) that can be probed may be ultimately limited by the instruments ability to measure the time interval $\Delta t$ between consecutive $\gamma$-ray flares.}. This is relatively moderate compared to what is achieved at other wavelengths (the ACIS instrument on-board the Chandra X-ray satellite achieves a resolution of $0.5'' \equiv \unit[20]{mpc}$, and near-infrared observations reach $(75\times 10^{-3})'' \equiv \unit[3]{mpc}$), and hampers the identification of GC $\gamma$-ray sources by comparing their position and morphology to possible counterparts observed at longer wavelengths.

\subsection{Large-scale morphology of the Central Molecular Zone}

\begin{figure}
\includegraphics[width=0.48\textwidth]{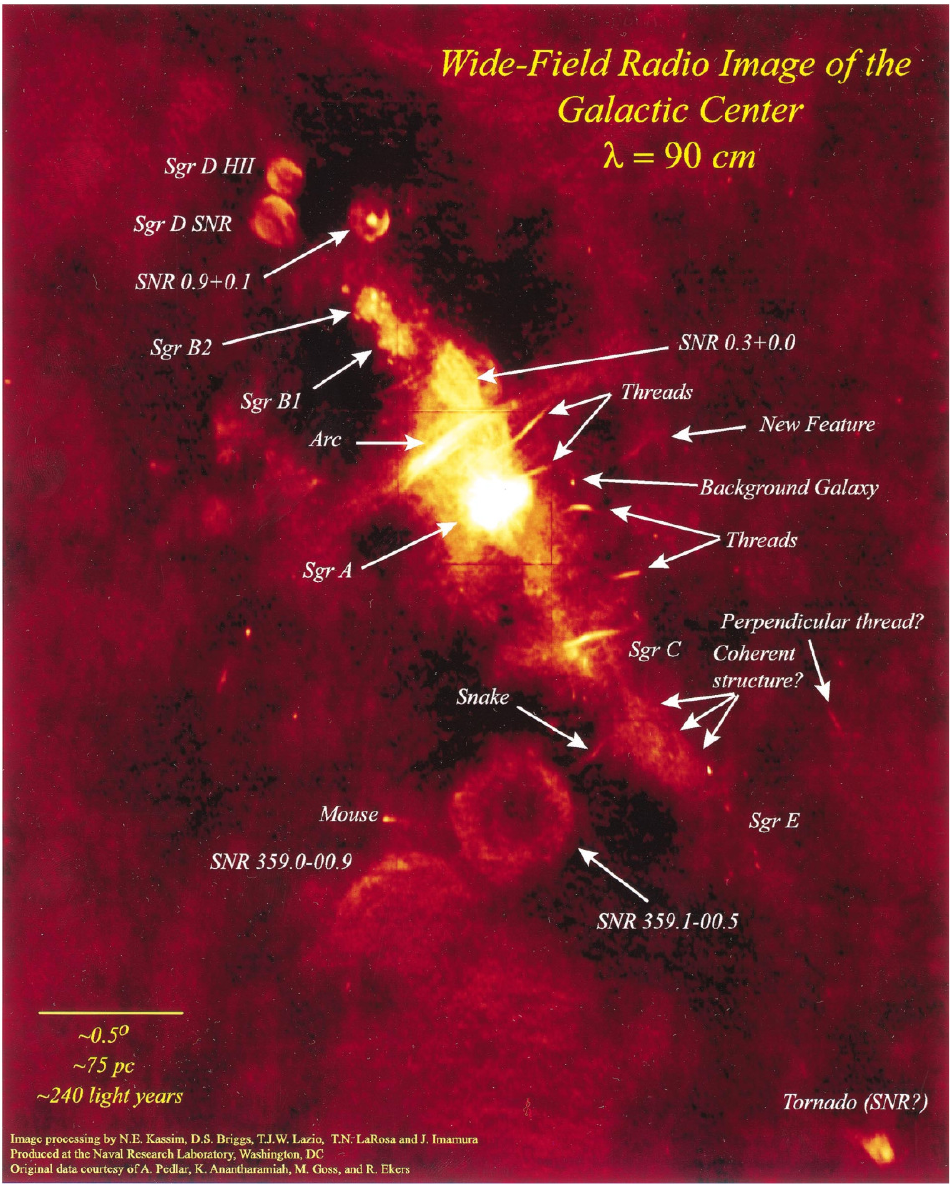}
\caption{Large-scale compilation of VLA 90~cm radio observations of
  the Galactic Centre region \cite{LaRosa00}. The distance scale is given on the bottom-left. The projection is in equatorial coordinates, such that the Galactic Plane is oriented top-left to bottom-right. Various SNRs and PWNe are labelled, which emit non-thermal synchrotron radiation caused by relativistic electrons. Several thread-like filaments, notably the so-called \emph{arc}, radiate synchrotron emission as well. The GC itself is located inside the Sgr~A complex, which is the brightest in this image. The full region, from the Sgr D molecular cloud in the east up to the radio source Sgr E in the west, has a diameter of about \unit[300]{pc} and is known as the Central Molecular Zone.}
\label{fig:LaRosa}
\end{figure}

Fig.~\ref{fig:LaRosa} shows the first large-scale compilation \cite{LaRosa00} of radio observations successively performed towards the GC with the VLA radio array at a wavelength of \unit[90]{cm}. The image reveals a complicated morphology due to the presence of numerous emission regions (that can at least in parts be identified with sites of particle acceleration), possibly interacting with one another. The emission is largely dominated by non-thermal synchrotron radiation, suggesting that acceleration of electrons to supra-thermal energies takes place throughout the region, possibly to energies of a few \unit[10]{TeV} and beyond. Several SNRs are clearly identified, and the PWN inside the SNR G~0.9+0.1 is well visible. The regions denoted by Sgr~B1, Sgr~B2, Sgr~C and Sgr~D contain large concentrations of ionised or molecular material, with gas densities of more than $\unit[10^4]{cm^{-3}}$, exceeding by far the density of clouds at other locations in the Galaxy. Several thread-like filaments, notably the GC arc, are oriented perpendicular to the Galactic plane and exhibit highly polarised radio emission with no line emission \cite{Yusef84}. The GC itself is located within the complex radio structure Sgr~A. On large scales, emission is concentrated along the Galactic plane, with an extension of about \unit[300]{pc} in Galactic longitude, and is known as the Central Molecular Zone (CMZ, \cite{Morris:1996}).

\begin{figure}
\includegraphics[width=0.48\textwidth]{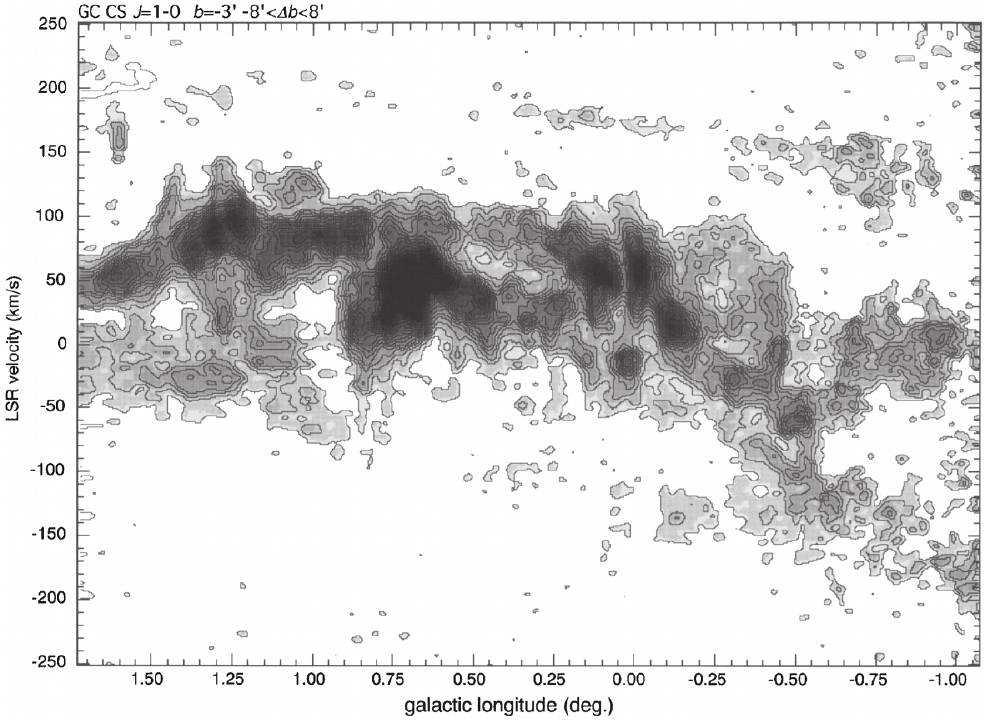}
\caption{Longitude-velocity diagram of CS radio emission from the CMZ, integrated over the latitude range $-11' \leq b \leq 5'$. The largest intensity, corresponding to the densest clouds, is seen east of the GC at positive velocities, while only much weaker emission is recorded from the region west of the GC. Image taken from \cite{Tsuboi:1999}.}
\label{fig:CS}
\end{figure}

The structure of molecular clouds in the region has been mapped
already in the 1970s using $^{12}\mathrm{CO}$ and $^{13}\mathrm{CO}$
lines \cite{bania77, liszt77}. Such measurements, however,
suffer from background and foreground
contamination from molecular clouds in the Galactic disk. In the
velocity range of interest for mapping the inner Galactic region, $|v|\lesssim\unit[100]{km/s}$,
carbon monosulfide (CS~$J=1\rightarrow 0$) line emission is expected to be essentially free of such
contaminations. Albeit being less sensitive because of its
larger critical density, CS emission provides an efficient tracer of
at least dense molecular clouds. The most complete CS survey
of the region is provided by measurements with the NRO radio telescope
\cite{Tsuboi:1999} and yields a total mass in molecular clouds of $(3-8)\times
10^7 M_{\astrosun}$ in the CMZ, which is about 10\% of the total molecular material of the Galaxy, and roughly consistent with earlier findings using different tracers (e.g.\ \cite{Damen:1998}). Fig.~\ref{fig:CS} shows a CS line emission intensity map of the CMZ, integrated over a range in Galactic latitude of $-11' \leq b \leq 5'$. A clear asymmetry in the distribution is observed, with the clouds east of the GC being much denser and moving (in projection) away from the observer, while the clouds west of the GC are less dense and move towards us.

Since the observability of line emission of a particular molecule depends on many parameters (such as molecule density, excitation energy, but also absorption along the line-of-sight), it is advantageous to trace the emission from as many different types of molecules as possible to get a consistent  view of the gas content in the CMZ. This is the idea of the MOPRA CMZ survey \cite{Jones:2012}, in which line emission from a large number of (complex) molecules was measured, complementing earlier works.

\begin{figure}
\includegraphics[width=0.3\textwidth]{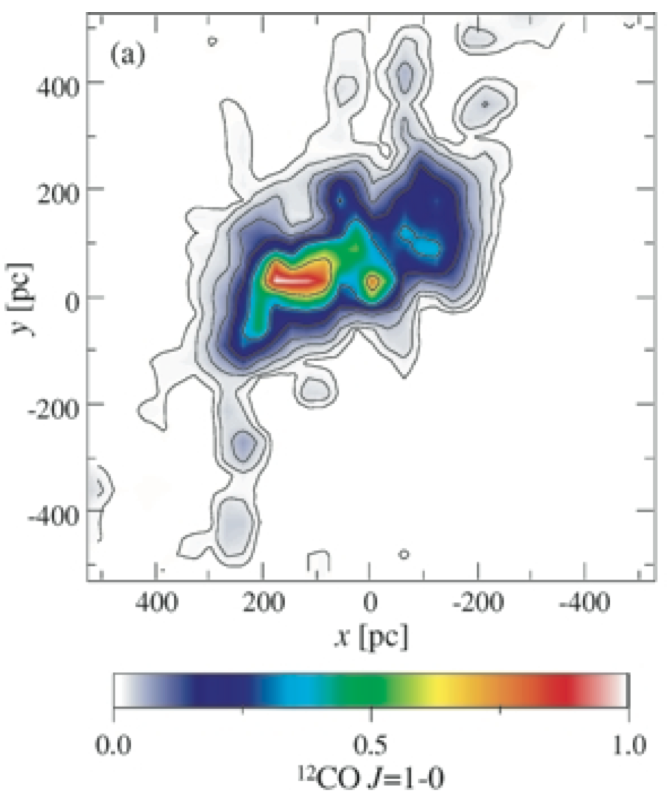}
\caption{Density of CO emission in a face-on-view, derived by comparing the CO line emission intensity of the gas to absorption of OH lines. The x-coordinate is oriented along the Galactic longitude, the y-coordinate denotes the line-of-sight distance w.r.t.\ the GC. The gas distribution mapped by the CO emission shows a ridge-like, elongated morphology, with the eastern part of the distribution being closer to the observer. Most of the gas is concentrated in a line-of-sight distance of $|y|\leq\unit[200]{pc}$. Regions of maximum intensity are approximately coincident with the regions of most intense CS emission (at Galactic latitude $l\sim 0.7^\circ$ and $l\sim 1.3^\circ$, see Fig.\ \ref{fig:CS}). Image taken from \cite{Sawada:2004}.}
\label{fig:CMZClouds}
\end{figure}

Based on either models of gas kinematics in the inner Galaxy \cite{Sofue:1995} or OH absorption measurements \cite{Sawada:2004}, attempts have been made to produce a face-on map of the CMZ gas distribution, i.e.\ to translate the longitude-latitude-velocity diagram (cf.\ Fig.~\ref{fig:CS}) into three-dimensional spatial distributions of gas density. These works are an important ingredient to study the gas dynamics in the CMZ, and, more important for the high-energy astrophysics discussed in this review, can serve as input to understand the morphology of diffuse $\gamma$-ray emission from modelling the local cosmic ray transport in that region. The studies suggest that the bulk of the gas content is located within a line-of-sight distance of $|z|<\unit[200]{pc}$ from the GC, with the eastern part containing the Sgr~B complex being located in front, and the western part behind the GC position. As an example, Fig.\ \ref{fig:CMZClouds} shows the face-on view gas distribution derived in a model-independent way by OH absorption studies \cite{Sawada:2004}.

Observations at far IR and submillimetre wavelengths provide a tracer of dust in the CMZ, which itself traces the molecular gas content. An analysis of Herschel IR data \cite{Molinari:2011} recorded at $\unit[70-500]{\mu m}$ wavelength suggests that the large-scale morphology of molecular gas in the CMZ has a roughly elliptically-twisted shape, with a total mass of about $\unit[3\times 10^7]{M_{\astrosun}}$, compatible with line emission measurements. About the same mass content ($\unit[5.3\times 10^{7}]{M_{\astrosun}}$) is derived from a SCUBA submillimetre survey of the CMZ \cite{Pierce:2000}. Besides that, IR observations are an ideal tool to study sites of star formation in the CMZ due to the excellent spatial resolution of these surveys and the low extinction of IR emission along the line-of-sight (among others, using the Spitzer space telescope \cite{Ramirez:2008,Hinz:2009,Yusef:2009}).

\begin{figure*}
\includegraphics[width=\textwidth]{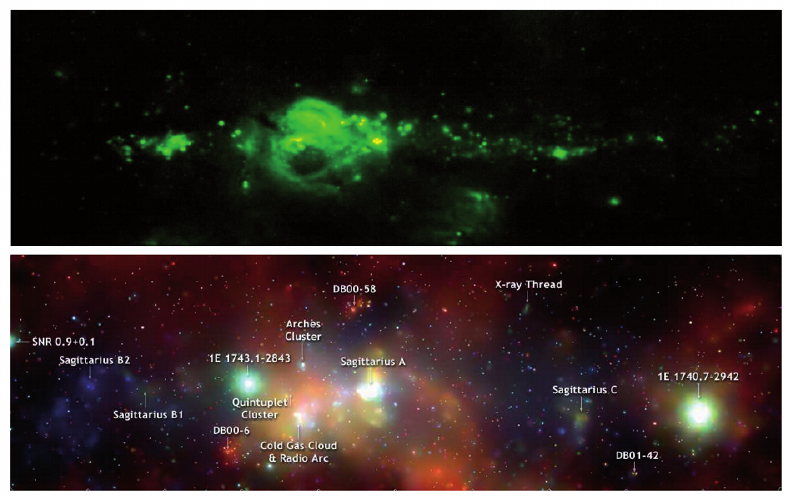}
\caption{Spitzer infrared (top) and Chandra X-ray (bottom) maps (in Galactic coordinates) of a $2^\circ\times 0.8^\circ$ region around the GC. The infrared emission traces warm dust and stellar activity, while the keV photons (red: \unit[1-3]{keV}, green: \unit[3-5]{keV}, blue: \unit[5-8]{keV}) trace high energy phenomena of the start and end points of stellar life (such as particle acceleration at stellar wind shocks or in large magnetic fields of rotating neutron stars) and extended regions where relativistic electrons produce synchrotron or bremstrahlung radiation. Images after \cite{Ramirez:2008} and \cite{Muno:2009}, respectively.}
\label{fig:SpitzerChandra}
\end{figure*}

Finally, a high resolution X-ray survey conducted with the ACIS instrument onboard the Chandra satellite provides a unique view onto the CMZ at photon energies of about $\unit[1-8]{keV}$ \cite{Muno:2009}. This study revealed the presence of more than 9000 X-ray sources, tracing mostly the high-energy phenomena connected to the stellar populations of the region, but also non-thermal X-ray sources such as PWNe and SNRs. Fig.\ \ref{fig:SpitzerChandra} shows Spitzer \cite{Ramirez:2008} and Chandra \cite{Muno:2009} maps of the inner $2^\circ\times 0.8^\circ$ region around the GC.

\subsection{The inner \unit[50]{pc} region}
The radio view of the inner 50~pc region is dominated by the Sgr A radio
complex (cf.\ Fig.~\ref{fig:LaRosa}), which can be substructured \cite{Downes:1971} into (i) the bright compact radio source \astar\ \cite{Balick:1974} at the dynamical centre of the Galaxy, (ii) the extended, non-thermal source \aeast\ (\cite{Ekers:1983}, 
at a projected distance of \unit[2.5]{pc} from the centre), enclosing in projection (iii) Sgr~A~West, a three-armed H~II region (the \emph{minispiral}) which rotates around the GC and exhibits a thermal radio spectrum. Sgr~A~West itself is surrounded by the so-called (iv) Circum Nuclear Disk of molecular gas of mass $\sim 10^4 M_{\astrosun}$ \cite{Mezger:1996}.

Based on X-ray \cite{Maeda2002,Park:2005,Koyama:2007} and radio
\cite{Jones74,Ekers:1983} observations, \aeast\ is nowadays identified as the remnant of a
supernova event which took place about 10.000 years ago (SNR~000.0+00.0). Since Galactic SNRs are established emitters of GeV and TeV $\gamma$-rays, \aeast\ is a good candidate counterpart for the compact HE and VHE emission observed coincident with the Sgr~A region (section \ref{sec:AEast}).
Spectral analysis reveals an overabundance of heavy elements in this object, favouring a supernova type II explosion of a 13-20~$M_{\astrosun}$ progenitor star. This interpretation is supported by the presence of a pointlike and offset X-ray source (\cite{Park:2005}, the \emph{cannonball}) which is probably the neutron star left over from the supernova explosion. \aeast\ itself shows a rather compact morphology compared to other Galactic SNRs, a natural explanation being that its shell expands into a region of dense (density $n \sim 10^3$~cm$^{-3}$) interstellar material, preventing a fast evolution of the forward shock \cite{Maeda2002}. Furthermore, the SNR is in contact with a dense molecular cloud ($n \sim 10^{5}$~cm$^{-3}$) on its eastern side. The radius of the X-ray emitting region of the remnant is even more compact than the one inferred from radio observations (\unit[2]{pc} vs.\ \unit[6-9]{pc}, respectively); while the radio emission traces the forward shock, the X-ray emission is thus probably caused by a reverse shock that heats plasma in the inner parts of the remnant. 

\subsection{The Galactic supermassive black hole and its immediate vicinity}
Since its discovery in 1974 \cite{Balick:1974}, the bright and ultra-compact radio source \astar, located at the dynamical centre of the Galaxy, is in the focus of GC research. Today, a wealth of precise astronomical observations support the idea that \astar\ is a supermassive BH. Since it is known that the dynamics of (active) galactic nuclei are largely driven by the presence of BHs in their cores, the GC offers the unique possibility to study in close view processes that are presumably at work in a large class of extragalactic nuclei as well (with the caveat that the energy output of the latter is much larger).

Key to establish the BH nature of \astar\ are observations with modern telescopes providing intrinsic resolution up to sub-milli-arcseconds. One of the most impressive studies is based on near IR observations of the orbits of young stars in the direct (as close as 0.1'' in projection) vicinity of \astar, from which the mass of the central compact object, $M_{\mathrm{A^*}} \sim 4\times 10^6~M_{\astrosun}$, can be inferred with great accuracy \cite{Schodel:2002,Ghez:2003,Eisenhauer:2005cv,Gillessen:2009}. At the same time, these studies show that the stars' orbits are consistent with a purely Keplarian motion around a point mass centred on the \astar\ radio position. These findings are supported by VLBA measurements \cite{Reid2004}, which put strong limits on the motion of the radio position of \astar\ itself w.r.t.\ the barycenter of the Galaxy as determined from stellar orbits. The results imply that the object responsible for the radio emission must be rather massive, and a lower limit of $~4\times 10^5~M_{\astrosun}$ on the mass of the \astar\ radio source is inferred from the measurements \cite{Reid2004}.
At a wavelength of \unit[7]{mm}, VLBI observations have resolved the size of the radio emission region to $24\pm 2$
Schwarzschild radii \cite{Bower2004}. Combining these findings, there is not much doubt that \astar\ can only be a supermassive BH
(see, e.g., the reviews \cite{Genzel:2007aa, Melia07} for more information).

The photon flux from the direction of \astar\ has been measured across a large range of energies. The energy spectrum in the millimeter to IR domain is
characterised by a hard power-law with photon index $\sim 0.3$, a turn-over at about \unit[1]{GHz}, followed by a cutoff at about $\unit[10^3]{GHz}$
\cite{Zylka1995}, explained as synchrotron radiation of relativistic electrons (e.g.\ \cite{Duschl1994, Melia2000}). 

While being relatively bright at radio frequencies, \astar\ is only a
faint X-ray emitter \cite{Skinner1987}, but shows bright outbursts on
time scales of a few minutes to several hours \cite{Baganoff2001, Porquet2003,Neilsen:2013,Degenaar:2013}. Assuming a black hole mass exceeding some $10^6$ solar masses and invoking causality arguments, these short flare durations limit the size of the emission region to be less than 10 BH Schwarzschild radii. Non-thermal processes near the event horizon might produce
relativistic electrons and thus explain the X-ray short-time variability (e.g.\ \cite{Markoff2001,Aharonian:2005ti,Liu2006}). To a certain extent, flares in the NIR band are predicted by these models, and such flares have been observed \cite{Genzel2003}.  Observations in the hard X-ray/soft \gr\ band by the INTEGRAL instrument, on the contrary, show a faint, but steady emission from the direction of the GC \cite{Belanger2006}.

At even higher energies, identification of any $\gamma$-ray emission with radio or X-ray counterparts is hampered by the comparatively modest angular resolution of HE $\gamma$-ray telescopes. Observations with the EGRET instrument onboard the Compton
Gamma-Ray Observatory in the late 1990's yielded a strong excess (named 3EG~J1746-2851) of
$>30$~MeV \grs\ on top of the expected Galactic diffuse
emission \cite{Hartmann:1999}. Within an error circle of $0.2^\circ$, the centroid of this
excess is compatible with the position of \astar. However, the energy output of 3EG~J1746
in the MeV-GeV range ($\sim 10^{37}$~erg~s$^{-1}$) exceeds by at
least an order of magnitude the energy released close to \astar\ at any other
wavelength. In any case, due to the relatively poor angular
resolution of EGRET, source confusion hampers the interpretation of
the signal especially at low energies, where the EGRET point spread
function is poorest. Furthermore, diffuse emission from the Galactic ridge is a dominating component at these energies, causing unavoidable systematic uncertainties in the position determination. A follow-up analysis of the position of 3EG~J1746,  using only events with energies $>1$~GeV to improve the instrument point spread function (PSF), disfavours an
association of the source with \astar\ at the 99.9\% CL \cite{Hooper2002}. New data taken with the Fermi-LAT instrument, however, significantly improve on these findings (see below).

Currently, \astar\ is in a rather low state of emission, since its bolometric luminosity is smaller than $10^{-8} L_{\mathrm{Edd}}$ (with the Eddington luminosity being $L_\mathrm{Edd}\sim \unit[10^{44}]{erg~s^{-1}}$ for a BH with \astar's mass). It is suggested that the BH is currently accreting only a moderate amount of gas from the winds of massive young stars populating the inner $\sim\unit[1]{pc}$ region \cite{Cuadra:2006}. This does not exclude that the GC was much more active in the past: indeed, there is evidence of recent ($\sim\unit[100]{years}$ ago) activity deduced from the presence of X-ray reflection nebulae \cite{Ponti:2010,Terrier:2010,Clavel:2013} in nearby molecular clouds, and even of times of much longer activity during the last $\unit[10^7]{years}$, as suggested by the presence of giant outflows from the GC region (e.g.~\cite{Su:2010}).

\section{Observing the sky at high and very high energies}
\label{sec:CurrentInstruments}

Due to the complex interplay between the different components present at the GC it is a challenging task to explain  in a consistent picture the emission from the region both on large and small scales. Key to a detailed understanding are (contemporaneous) multi-wavelength observations covering an as broad range in photon energy as possible, with good sensitivity to flares and an excellent angular resolution. Impressive progress in reaching this goal has been achieved during the last decades at $\lesssim \unit[10]{keV}$ energies (see section \ref{sec:stage}). Similar progress has been made in the detection of HE and VHE photons, where it is nowadays possible to produce well-resolved sky maps, light curves and energy spectra of sources. However, instruments detecting \unit{MeV-TeV} $\gamma$-rays lack a comparable angular resolution and flare sensitivity, making it hard to match their data with observations at lower energies. The reason for this is that (among others)
\begin{itemize}
\item the energy distributions of non-thermal particle populations generally feature the form of a power-law, $dn/dE\sim E^{-\Gamma}$, with  spectral indices $\Gamma\sim 2\dots 3$, meaning that only very few particles reach highest energies. As the resulting differential $\gamma$-ray spectrum exhibits a similar energy dependence, and the detection area of $\gamma$-ray instruments is limited, the number of detected photons is small, limiting the ability to detect flares and produce well-resolved energy spectra (especially at the high energy limit of the instruments).
\item in current detectors, high energy photons can only be detected rather indirectly through production of secondary particles, which limits both the energy and angular resolution.
\end{itemize}
On the other hand do HE and VHE photons provide a very direct tracer of sites of particle acceleration to highest energies and interaction of these particles with the surrounding medium (either the gas distribution or interstellar radiation fields), making them an indispensable element of any systematic multi-wavelength study of high energy processes in the universe.

The HE and VHE $\gamma$-ray flux arriving at Earth from typical $\gamma$-ray sources is low. As an example, from the direction of the GC, a fairly strong VHE $\gamma$-ray emitter, a flux well below $\unit[1]{photon~m^{-2}~yr^{-1}}$ is recorded for $\gamma$-rays with energies above \unit[1]{TeV}. This is the reason why VHE $\gamma$-ray instruments are ground-based: Imaging Atmospheric Cherenkov Telescopes (IACTs) use Earth's atmosphere as a detector medium, and record the Cherenkov light emitted by relativistic electrons and po\-si\-trons that are copiously produced when the $\gamma$-ray generates an air shower in the upper atmosphere (for recent reviews, see, e.g., \cite{Hinton:2009, Volk:2009}). This detection technique provides a large detection area ($\unit[10^4-10^5]{m^2}$, depending on the ground area covered by the instrument and the $\gamma$-ray energy). Given typical observation times of $\unit[1-100]{h}$, this guarantees good $\gamma$-ray statistics to study morphology and energy spectrum of the emission. The typical detection energy threshold is about $\unit[30-300]{GeV}$ and depends strongly on the zenith angle of the observation. Detection of VHE $\gamma$-rays suffers from background caused by the much more numerous charged cosmic rays impinging Earth's atmosphere. This background can, however, be efficiently suppressed in modern instruments. During the last ten years, this new generation of IACTs opened up a new observational window to the universe: more than 150 VHE $\gamma$-ray sources -- both Galactic and extragalactic -- have been discovered since then \cite{TeVCAT}, and at least 6 source classes identified \citep{Hinton:2009}. Observations with the currently operated systems H.E.S.S. (located in the southern hemisphere) and MAGIC and VERITAS (in the northern hemisphere) cover the entire sky, with the H.E.S.S.\ instrument suited best for GC observations, because at the location of the H.E.S.S.\ site, the GC region culminates close to zenith. The typical field-of-view (FOV) of these instruments is about $3-5^\circ$ in diameter, i.e.\ the instruments must be pointed at regions of interest and cannot be used easily for all-sky surveys. This drawback is avoided by water-Cherenkov instruments like e.g.\ HAWC \cite{HAWC:2013} or ARGO-YBJ \cite{Aielli:2007}, which monitor large portions of the sky at the expense of a reduced $\gamma$-ray sensitivity and a worse angular as well as energy resolution, and are particularly well-suited for transient searches. Current water-Cherenkov instruments are located only in the northern hemisphere, with (almost) no access to the GC region.

\begin{figure}
\includegraphics[width=0.48\textwidth]{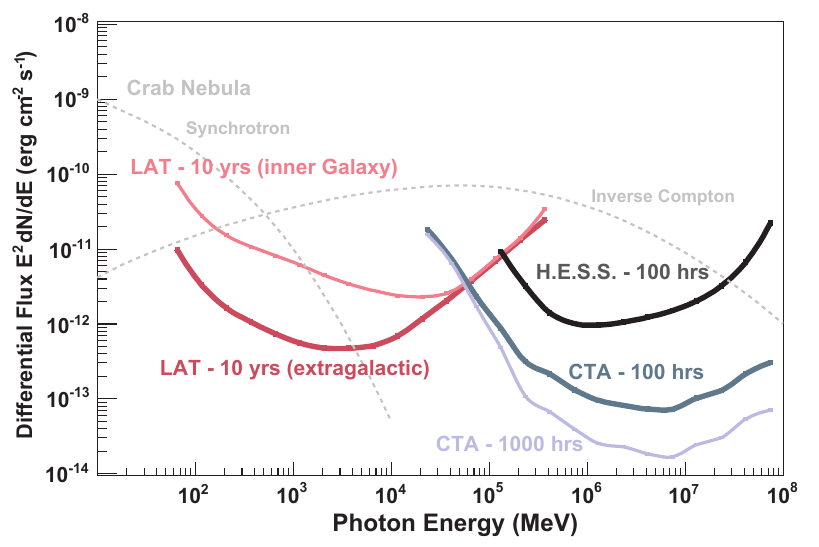}
\caption{Comparison of differential sensitivities for Fermi-LAT and H.E.S.S.\ for point-like $\gamma$-ray sources. Shown is the minimum differential flux (multiplied by the energy squared) that can be detected with a statistical significance of 5 standard deviations above the background (additionally assuming a systematic uncertainty of 5\% on the accuracy of the background estimate). The differential energy spectra of synchrotron and IC radiation from the Crab nebula (a bright, point-like $\gamma$-ray source) is shown for comparison. While Fermi-LAT reaches a sensitivity of about $\unit[10^{-12}]{erg~cm^{-2}~s^{-1}}$ at an energy of $\sim\unit[1]{GeV}$ in 10 years of operation, H.E.S.S. reaches about the same sensitivity at $\sim\unit[1]{TeV}$ in about \unit[100]{h} observation time. The expected sensitivity of the planned CTA observatory is given for comparison. Since diffuse $\gamma$-ray background plays an important role, the LAT sensitivity in the Galactic Plane is worse than that for extragalactic observations. Note that the LAT sensitivity is reached for every point in the sky, whereas the H.E.S.S.\ sensitivity is for an observation of a single region of interest. Towards larger energies, sensitivity degrades because of lack of photon statistics. At the low energy end, the sensitivity is limited by background systematics. Figure taken from \cite{Funk:2013}.}
\label{fig:Sensitivities}
\end{figure}

HE $\gamma$-rays are detected by satellite-born instruments. Due to the limiting payload, these detectors are small and feature a detection area of $\lesssim \unit[1]{m^2}$. However, the typical $\gamma$-ray flux in the core energy range of these detectors (\unit[10]{MeV}--\unit[10]{GeV}) is much larger than at VHE energies, such that the small area is sufficient to obtain good photon statistics. The photons are detected via electron-positron pair-production in converter material, and the paths of the charged particles is reconstructed by position-sensitive tracking detectors, followed by an electromagnetic calorimeter to estimate the energy of the primary $\gamma$-ray. The current missions Fermi-LAT \cite{LAT:2009} and AGILE \cite{Tavani:2009} improve upon the predecessor EGRET in that they provide a larger sensitive detector area at better angular and energy resolution. Fermi-LAT plays an important role in understanding both the Galactic and extragalactic high-energy universe. In its second source catalogue, 1873 HE $\gamma$-ray sources are listed \cite{Nolan:2011}, which is a factor 10 improvement over EGRET just in terms of source statistics. HE $\gamma$-ray detection is almost free of cosmic ray background thanks to veto detectors installed at the outer surface of the instrument. Identification of sources close to the Galactic plane is instead hampered by the presence of diffuse $\gamma$-ray emission \cite{Ackermann:2012} which often has to be adequately modelled and subtracted to study individual sources (at VHE energies, diffuse emission is suppressed because its energy spectrum, compared to that of individual sources, falls off more steeply as energy increases). As opposed to IACTs, satellite-born detectors have a rather large field-of-view and can monitor the entire sky several times per day. Flux sensitivites of the LAT and the H.E.S.S.\ instrument are compared in Fig.~\ref{fig:Sensitivities}. 
 
 \section{Part I. GeV and TeV gamma-ray astrophysics of the Galactic Centre region}
\label{sec:part1}
\subsection{A strong TeV point source at the Galactic Centre}
\label{sec:Discovery}

Given the importance of the GC as a possible multi-TeV particle accelerator, the region was in the focus of IACTs since the early days of this detection technique (which dates back to the successful start of the Whipple single-dish telescope in 1968). It took, however, until 2004 that a VHE $\gamma$-ray signal was detected from the GC by the Whipple, Cangaroo-II and H.E.S.S.\  instruments almost simultaneously \cite{Kosack:2004ri,Tsuchiya:2004wv,Aharonian:2004wa}. The collaborations reported a point-like emission from the direction of \astar\ with no hint for flux variability. For the purpose of this review, this VHE $\gamma$-ray source will be called HESS~J1745-290.

Due to its location in the northern hemisphere, the Whipple instrument observed the GC only at rather large zenith angles (and hence a large energy threshold). It reported detection of the GC source with a marginal significance of 3.7 standard deviations above the background, and published a $\gamma$-ray flux of $(1.6\pm 0.5_{\mathrm{stat}} \pm
0.3_{\mathrm{sys}})\times 10^{-12}$~cm$^{-2}$~s$^{-1}$, integrated above an energy threshold of \unit[2.8]{TeV} \cite{Kosack:2004ri}. The Cangaroo-II instrument in Australia, instead, reported a significant detection of the source above a photon energy of \unit[250]{GeV} and provided a measurement of its energy spectrum which showed a steep decline with energy (power law in energy with photon index of $4.6\pm 0.5$) \citep{Tsuchiya:2004wv}. Observations during 2003 with two telescopes of the partially completed H.E.S.S. array also resulted in a clear detection of \hgc\ \citep{Aharonian:2004wa} above an energy threshold of \unit[165]{GeV}. As opposed to Cangaroo-II, a hard energy spectum (photon index of $2.20\pm 0.09_{\mathrm{stat}}\pm 0.15_\mathrm{sys} $) of the source was reported. These results were later confirmed by follow-up observations carried out in 2004 with the completed H.E.S.S. array \cite{Aharonian:2006wh} and by observations of the MAGIC instrument \citep{Albert:2005kh}. Recently, the VERITAS collaboration published an independent measurement of the $\gamma$-ray spectrum (\cite{VERITAS:2014}, above a threshold energy of \unit[2.5]{TeV}) which matches the H.E.S.S.\ and MAGIC results both in index and flux normalisation, such that the initial disagreements about the spectral properties of HESS~J1745-290 can be regarded as settled.

\begin{figure}
\includegraphics[width=0.48\textwidth]{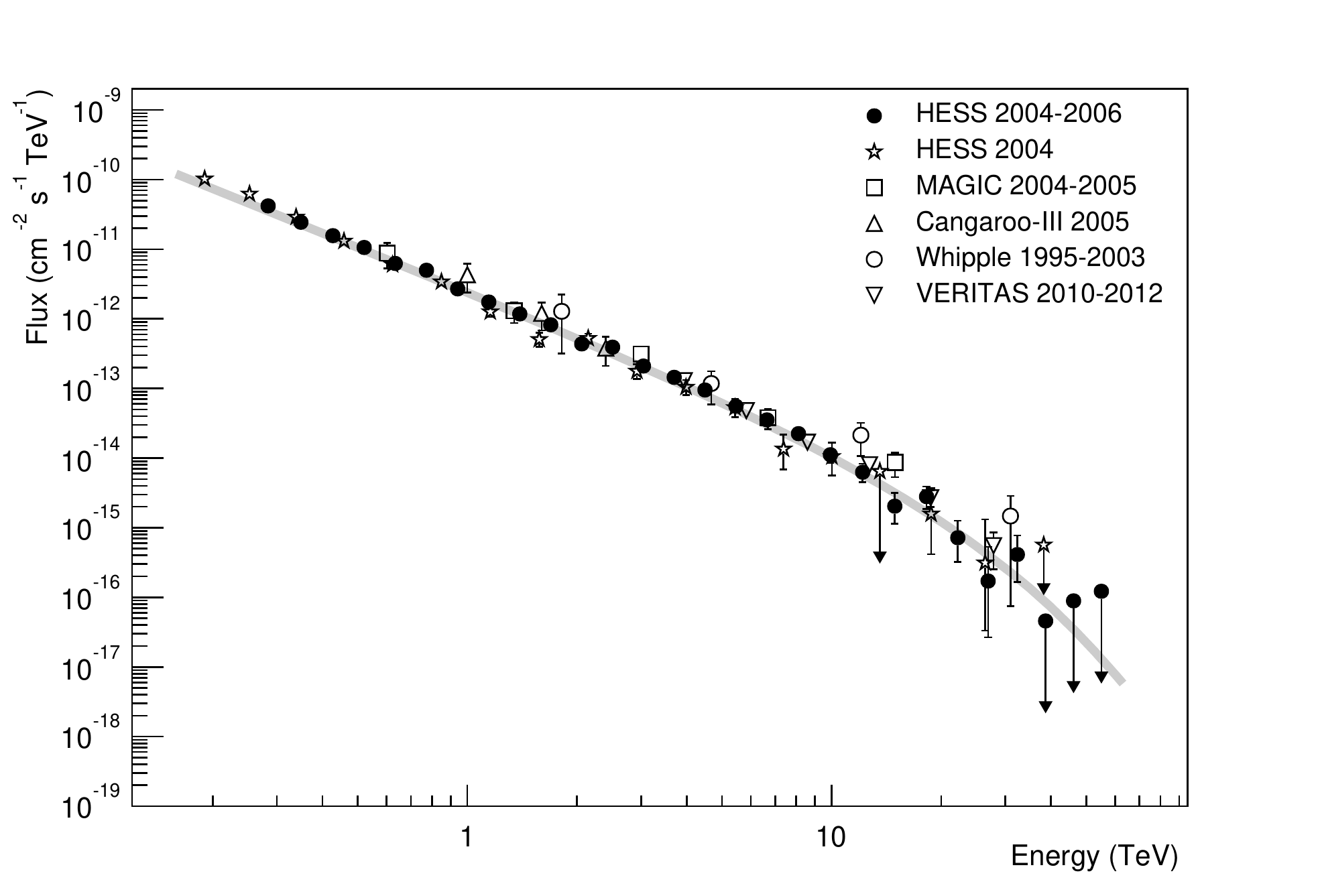}
\caption{Compilation of VHE $\gamma$-ray spectra
of the GC source HESS~J1745-290. Data points are
taken from \cite{Kosack:2005, Aharonian:2006wh, Albert:2005kh, Mizukami2008, GCSpectrum, VERITAS:2014}. 
The curve shows a power-law fit with exponential cut-off to the most recent H.E.S.S. data \cite{GCSpectrum}. Note that the H.E.S.S. spectra were corrected for a flux contribution of $\sim 15$\% from diffuse emission. Upper limits are given at 95\% CL. The early results from Whipple \citep{Kosack:2004ri} and Cangaroo-II \citep{Tsuchiya:2004wv} are not shown.}
\label{fig:Spectra}
\end{figure}

Due to its location in the southern hemisphere, the H.E.S.S.\ instrument currently has the most sensitive view onto the GC region, yielding both a good photon statistics at the highest energies of the spectrum and a low energy threshold, making its data well suited for spectral studies. Based on 93 hours (live time) of observations during the years 2004-2006, for the first time a clear deviation of the energy spectrum from a single power-law distribution was observed: the spectrum is well described by a power-law with exponential cut-off,
\begin{equation*}
\frac{dN}{dE} = \Phi_0 \cdot \left( \frac{E}{1\mathrm{TeV}} \right)^{-\Gamma} e^{-\frac{E}{E_\mathrm{c}}},
\end{equation*}
with $\Phi_0 = (2.40\pm 0.10)\times 10^{-12}$~TeV$^{-1}$cm$^{-2}$s$^{-1}$, $\Gamma=2.10\pm 0.04$, and $E_\mathrm{c}=\unit[(14.70\pm 3.41)]{TeV}$ \cite{GCSpectrum}. The data are equally well described by a smoothed broken power-law, but a simple power-law fit is clearly rejected. Fig. \ref{fig:Spectra} shows a compilation of the at date available VHE $\gamma$-ray flux measurements of HESS~J1745-290, together with the above given fit to the latest H.E.S.S.\ data, indicating the recent agreement between the different instruments. A recent (yet preliminary) update of the H.E.S.S.\ flux spectrum \cite{Moulin:2013} suggests that the source-intrinsic energy cut-off moves to lower values, $E_\mathrm{c}\sim \unit[7]{TeV}$, when correcting the spectrum for the underlying diffuse $\gamma$-ray emission.

\subsection{The MeV/GeV Galactic Centre source}
\label{sec:HESource}

\begin{figure}
\includegraphics[width=0.48\textwidth]{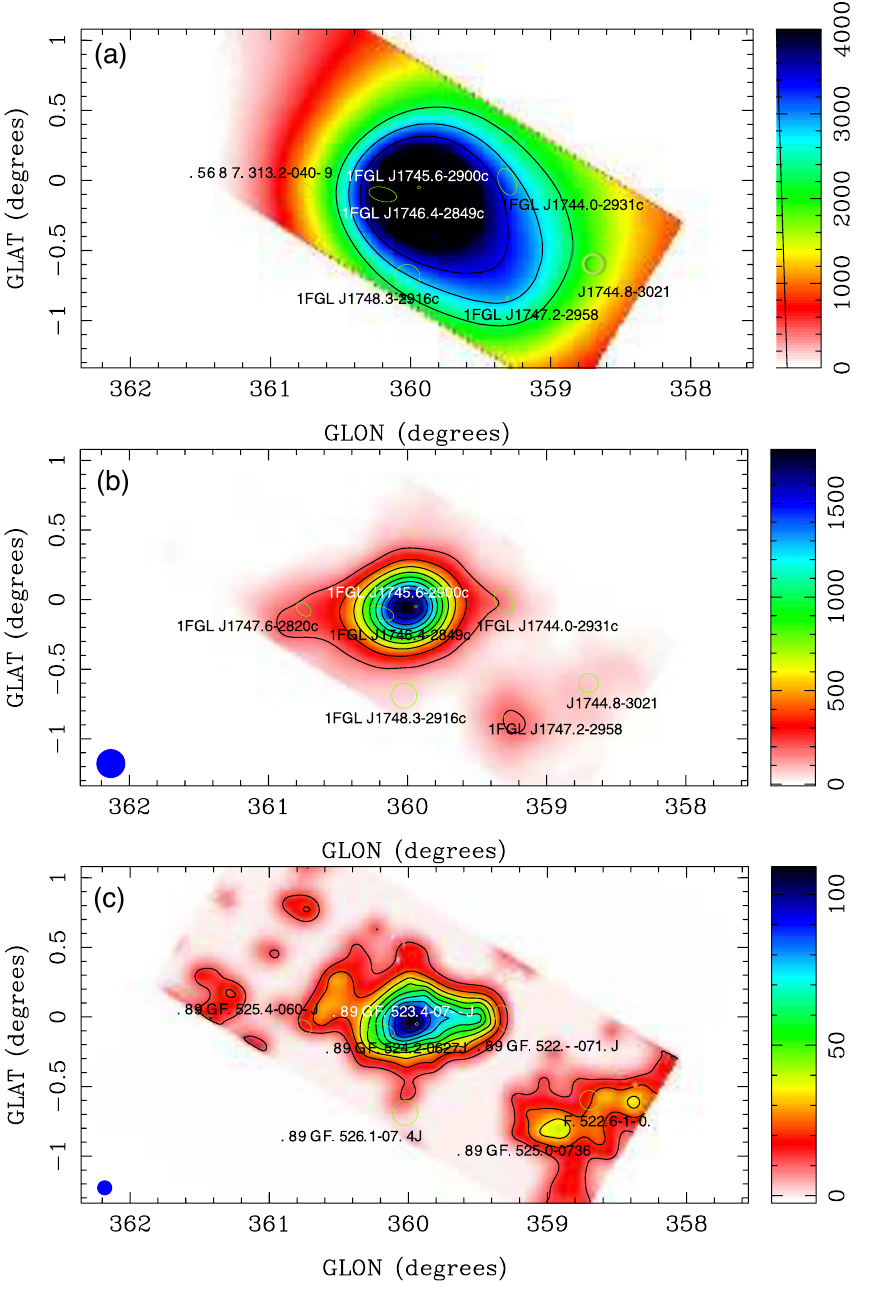}
\caption{TS maps (\emph{test statistics}, roughly corresponding to the square of the statistical significance in units of standard deviations) of the CMZ as seen by the Fermi-LAT instruments after 25 month of data taking \cite{Chernyakova:2011}, for photon energy intervals \unit[0.3-3]{GeV}, \unit[3-30]{GeV}, and \unit[30-300]{GeV} (top, middle, bottom, respectively). The angular resolution of the instrument is shown as a circle in the lower-left corner of each map. The position error circles of six identified HE $\gamma$-ray sources listed in the 1FGL catalogue of Fermi-LAT $\gamma$-ray sources \cite{Abdo:2010} are shown by green ellipses, among them the source 1FGL~J1745.6-2900, which is, according to the second Fermi-LAT source catalogue \cite{Nolan:2011}, possibly associated with \aeast\ (called 2FGL~J1745.6-2858 in this catalogue).}
\label{fig:FermiGC}
\end{figure}

\begin{figure}
\includegraphics[width=0.48\textwidth]{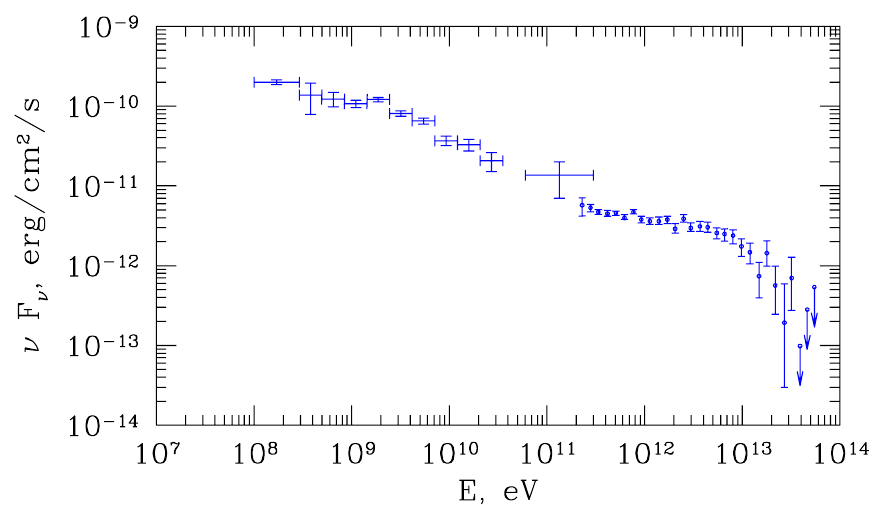}
\caption{Spectral energy distribution of the GC point source. Data points up to an energy of $\sim\unit[100]{GeV}$ are from the Fermi-LAT source 2FGL~J1745.6-2858 \cite{Chernyakova:2011}, and data points above that energy from H.E.S.S.\ \cite{GCSpectrum}. Despite known caveats in flux extraction (different flux integration regions due to different angular resolutions of the instruments, systematic uncertainties due to diffuse emission and hadronic background), the energy spectra match well, suggesting a common origin of the emission. Figure after \cite{Chernyakova:2011}.}
\label{Fig:GCCombinedSpectrum}
\end{figure}

As mentioned earlier, diffuse $\gamma$-ray emission is much more pronounced high energies, making the identification of individual sources much more difficult than at very high energies. As can be seen from Fig.~\ref{fig:FermiGC} which is based on 25 month of Fermi-LAT data of the region \cite{Chernyakova:2011}, the relative intensity of the diffuse emission is strongly reduced, as well as the angular resolution improved, when the energy is increased, meaning that source identification is generally easier using a high-energy selection of the data (but at the expense of photon statistics). Coincident with the position of the GC, a hard HE $\gamma$-ray source (named 2FGL~J1745.6-2858 in the second source catalog of Fermi-LAT sources \cite{Nolan:2011} and denoted 1FGL~J1745.6-2900 in Fig.~\ref{fig:FermiGC}) is visible. By performing a global fit to the $\gamma$-ray count map (including diffuse emission as well as known point sources), the $\gamma$-ray spectrum of 2FGL~J1745.6-2858 was derived in the energy range \unit[0.3--100]{GeV} \cite{Chernyakova:2011}. The spectrum is best described by a broken powerlaw with spectral indexes $\Gamma=2.20\pm 0.04$ and $\Gamma=2.68\pm 0.05$ below and above, respectively, a break energy of $E_\mathrm{b} = \unit[2]{GeV}$. No hint for flux variability of this source is found. Although the flux spectrum at its high energy end is significantly steeper than that of \hgc\ ($\Gamma = 2.1$, sect.~\ref{sec:Discovery}), the overall fluxes of the two sources match well (see Fig.~\ref{Fig:GCCombinedSpectrum}), suggesting a common origin of the HE and VHE emission. Due to the complicated spectral shape, an identification of the emission process is particularly difficult. Models that aim to describe the emission in the whole energy range are still relatively rare (but see \cite{Chernyakova:2011,Linden:2012a} who suggest that the emission may well be produced by proton-gas interactions in the immediate vicinity of the accelerator), as the HE source was only detected recently. Instead many models concentrate on the VHE emission processes. In this review, we will therefore summarise primarily emission models that are put forward to explain the VHE emission of \hgc, and connect to the HE emission of 2FGL~J1745.6-2858 when appropriate.

\subsection{The nature of the central accelerator}
\label{sec:HGC}
Despite the recent progress in obtaining a consistent picture of the GC $\gamma$-ray emission, the actual mechanism that produces the point-like emission is not yet understood. As discussed in section \ref{sec:CurrentInstruments}, a firm identification is particularly hampered by the --
compared to radio or X-ray instruments -- modest angular resolution of
the current generation of instruments. Compared to the projected distances of counterpart candidates in the GC, the $\gamma$-ray emission covers a  relatively large region, giving rise to source confusion in this densely populated part of the Galaxy. In particular, it is not entirely clear if the $\gamma$-ray emission measured by Fermi-LAT and ground-based Cherenkov instruments are produced by the same astrophysical object. Assuming that this is the case, the $\gamma$-ray picture of the source must fit into the wealth of GC observations at other wavelength (which each feature a different FOV and angular resolution). Nevertheless $\gamma$-ray observations can constrain counterparts and emission models in various ways. Without being in conflict with measurements at other wavelengths, models for HESS~J1745-290 and 2FGL~J1745.6-2858 must explain (at least) the following source properties \cite{GCSpectrum,Chernyakova:2011,Acero:2010,VERITAS:2014}:
\begin{itemize}
\item The energy spectrum between \unit[100]{MeV} and \unit[30]{TeV} is in general characterised by a hard powerlaw in photon energy. At various energies ($\sim\unit[2]{GeV}$, $\sim\unit[20]{GeV}$, $\sim \unit[10]{TeV}$) the spectrum features spectral breaks and a cut-off (Fig.~\ref{Fig:GCCombinedSpectrum}).
\item There is no hint for flux variability on any time scale from minutes to years\footnote{It should be noted that due to the limited photons statistics the sensitivity to flare detection depends strongly on the flare duration. \cite{GCSpectrum} estimate that a factor 2 (0.3) increase w.r.t.\ the quiescent flux is needed to detect a flare of a few hours (90 days) duration.}.
\item The emission region of the H.E.S.S.\ source is point-like and coincident with the position of \astar. The intrinsic size of the VHE source amounts to less than 1.3 minutes of arc. There is a hint for a moderate extension of the Fermi-LAT source.
\end{itemize}

Although an association of the $\gamma$-ray emission with Sgr~A* is compelling (and certainly viable in terms of energetics, position and spectrum, see below), there are at least two other objects in direct vicinity of the BH which are good candidates for producing the observed $\gamma$-ray flux in parts or in total: the SNR \aeast\ and the recently discovered PWN G359.95-0.04 \cite{Wang:2005}. Besides that, the $\gamma$-ray emission could be produced by the cumulative effect of many sources, such as a large population of millisecond pulsars \cite{Bednarek:2013}. The most important arguments in favour or against the various possible associations are summarised below. In anticipating the results of the studies, we note that both HE and VHE $\gamma$-ray emission (on their own or in total) can be explained by a variety of different sources and emission mechanisms, and this  confusion prevents a firm identification of \hgc\ and 2FGL~J1745.6-2858 until today.

\subsubsection{Sgr~A~East as a candidate $\gamma$-ray source}
\label{sec:AEast}
The existence of synchrotron radiation, i.e.~the presence of relativistic electrons,
and a large magnetic field ($\sim \unit[2-4]{mG}$, \cite{Yusef96}) make \aeast\ an excellent
candidate $\gamma$-ray emitter. Indeed is the discovery of HE $\gamma$-ray emission by EGRET (3EG~J1746, see section~\ref{sec:stage}) in support of this idea, since the presence of HE $\gamma$-ray emission is by itself a clear indication for particle acceleration to at least an energy of \unit[100]{GeV}. Given possible source confusion due to the limited angular resolution of EGRET, the case for a one-to-one association of the HE $\gamma$-ray source and the SNR is however far from compelling. On the one hand is the spectral shape of HE $\gamma$-ray emission of 3EG~J1746 found similar to that of other EGRET SNRs. On the other hand does the total energy radiated in the HE domain exceed that observed for other EGRET SNRs by about two orders of magnitude, making it at first difficult to believe that the HE emission is produced by the SNR alone. However, \aeast\ is in contact with dense molecular material (see section~\ref{sec:stage}); therefore, it might be argued that the enhanced $\gamma$-ray luminosity is due to shock-accelerated protons interacting with the gas of the dense molecular cloud \cite{Fatuzzo2003}. Such a scenario can self-consistently account for both soft ($<\unit[100]{MeV}$) $\gamma$-ray and radio emission by bremsstrahlung processes and synchrotron emission, respectively, of relativistic electrons produced as secondary particles in the inelastic proton-gas interaction. For a \unit[4]{mG} magnetic field, the estimated maximum energy to which protons can get accelerated in the remnant is $\sim\unit[10^{19}]{eV}$ \cite{Crocker2005}, supporting the idea that even VHE emission should be observable from \aeast. If indeed the emission from \hgc\ was due to the SNR, the lack of VHE flux variability could be naturally explained by the fact that proton acceleration takes place in the extended shell of the SNR (coherent variablility on short time scales is restricted to small-sized regions by causality arguments).

\begin{figure}
\includegraphics[width=0.48\textwidth]{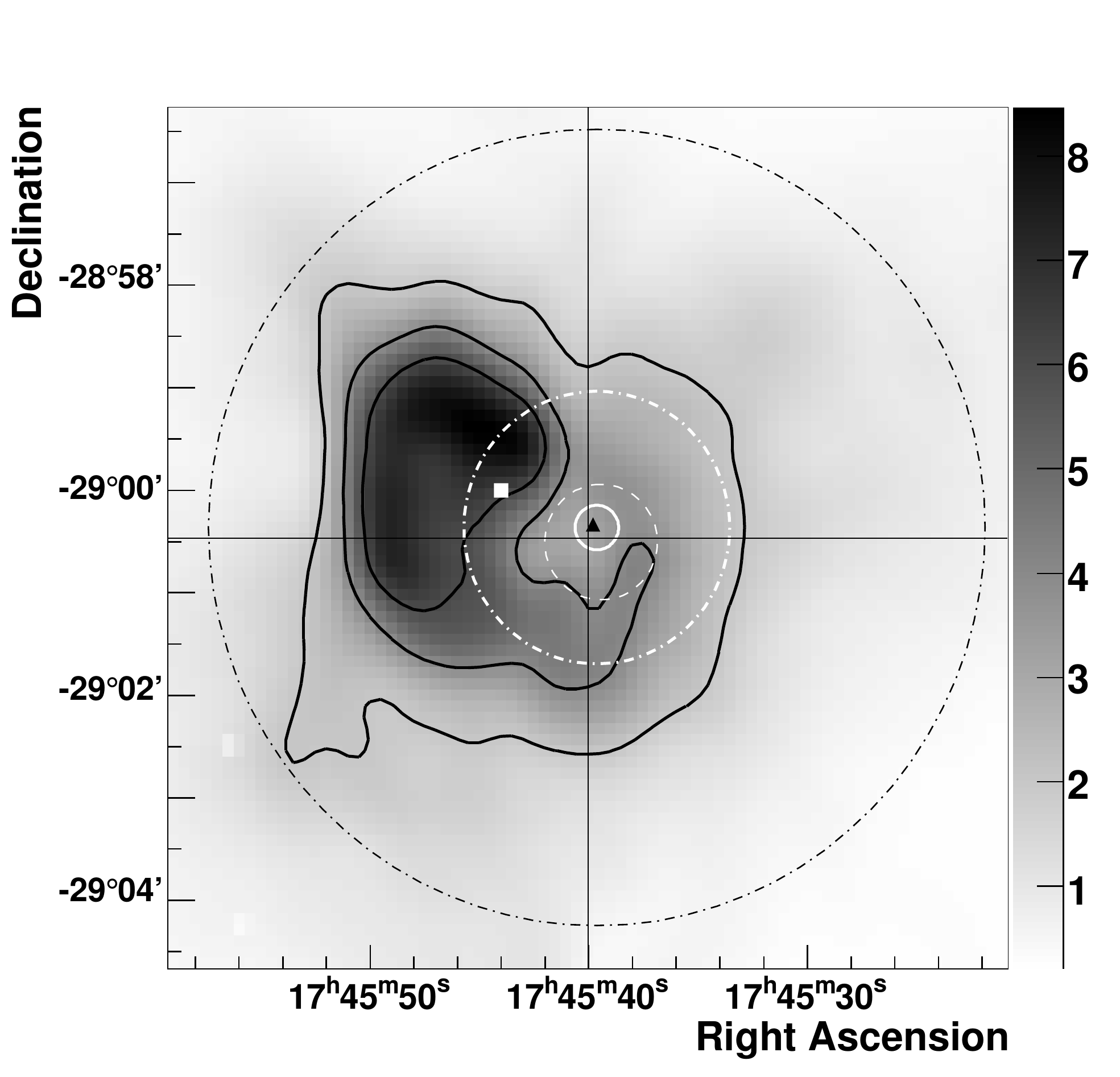}
\caption{90~cm VLA radio flux density map \citep{LaRosa00} of the
    innermost 20~pc of the GC, showing emission from the SNR \aeast.
    Black contours denote radio flux levels of 2, 4, and 6~Jy/beam.
    The centre of the SNR \citep{Green:2009} is marked by the white
    square, and the positions of \astar\ \citep{Reid:1999} and G395.95-0.04
    (head position, \cite{Wang:2005}) are given by the cross hairs
    and the black triangle, respectively. The 68\% CL total error contour of
    the best-fit centroid position of \hgc\ is given by the white
    circle. The dashed white circle shows the same contour for the
    previously reported H.E.S.S. measurement \cite{Aharonian:2006wh}.
    The white and black dashed-dotted lines show the 95\% CL upper limit
    contour of the $\gamma$-ray source extension and the 68\% containment region of
    the H.E.S.S.\ PSF, respectively.}
\label{fig:RadioMap}
\end{figure}

Despite the fact that the angular resolution of $\gamma$-ray instruments is poor, point-like sources can be localised rather precisely since the position uncertainty of their centroids roughly scales as $\theta/\sqrt{n}$, where $\theta$ and $n$ are the angular resolution and the number of detected $\gamma$-rays, respectively. Hence, for a strong point-like source of $\gamma$-rays and/or a long exposure, arcsecond precision on the source centroid location can be obtained (if systematic uncertainties are under control). Based on a precise measurement of the centroid of \hgc, 
Sgr~A~East is strongly disfavoured as the main counterpart of the VHE emission \cite{Acero:2010}. This is illustrated in Fig.~\ref{fig:RadioMap} which shows the \unit[90]{cm} radio flux of \aeast\ as a shell-like structure, surrounding \astar\ in projection. Note that the region is fully contained within the size of the H.E.S.S.\ point spread function, meaning that a per-$\gamma$-ray source identification is difficult. However, the centroid of \hgc, with a 68\% CL total error radius of 13'' only, is located in a region where the radio emission from the SNR is comparatively low. Therefore, it appears unlikely that VHE and radio emission are of the same origin. By means of a statistical analysis, \aeast\ can be ruled out as the bulk emitter of the VHE $\gamma$-rays at a significance of 3.9 standard deviations for the most conservative approach \cite{Acero:2010}. The position measurement is the most precise so far in VHE $\gamma$-ray astronomy, and was only achieved after a careful investigation of the pointing systematics of the H.E.S.S. telescopes, reducing the systematic error on the centroid position from 20'' \citep{Aharonian:2006wh} to 6'' per axis for this data set. No such strong conclusion can be drawn for the Fermi-LAT source yet, since (i) photon statistics is small compared to the VHE data set and (ii) systematic errors arising from the diffuse emission model are large.

\subsubsection{The Galactic Centre source: a pulsar wind nebula?}
\label{sec:PWN}
The recent detection of the PWN G359.95-0.04 in a deep Chandra
exposure of the GC region \citep{Wang:2005} complicates
the identification of the central particle accelerator even further. As G359.95-0.04 is located only $8.7''$ in
projection away from Sgr~A* (see Fig.~\ref{fig:RadioMap}), corresponding to a projected physical distance of \unit[0.3]{pc}, a discrimination between these two objects by
means of a position measurement of the VHE and HE $\gamma$-ray signal is impossible with current instruments.  With an implied luminosity of $\unit[10^{34}]{erg~s^{-1}}$ in the $\unit[2-10]{keV}$ band \cite{Wang:2005}, G359.95-0.04 is rather faint at X-ray energies, yet about four times brighter than \astar\ at these energies. The PWN shows a cometary shape and exhibits a hard and non-thermal X-ray spectrum which gradually softens when going away from the so-called head of the nebula, where the yet undiscovered pulsar is believed to be located. No radio counterpart of the PWN is found.

The fact that the X-ray spectra steepen with increasing distance from the pulsar position is an indication that the non-thermal electron population is cooled by synchrotron radiation. Using this information, a lower limit of $\unit[100]{\mu G}$ can be derived on the value of the magnetic field, assuming typical Galactic Centre radiation fields \cite{Hinton:2006zk}. Despite the fact that the high-energy electrons are rapidely cooled away, calculations show that a population of non-thermal electrons from the PWN can naturally account for both the X-ray photons from G359.95-0.04 and the VHE $\gamma$-ray emission of \hgc. The reason is that dense stellar radiation fields provide a rather unique environment for these electrons (compared to other locations in the galactic disk): in the scenario proposed by \cite{Hinton:2006zk}, far IR photons get up-scattered by the IC process to VHE energies. Because of the large abundance of these photons and despite strong synchrotron cooling, a small number of electrons suffice to provide the large $\gamma$-ray luminosity of \hgc\ (roughly an order of magnitude larger in the \unit[1--10]{TeV} energy band than in the \unit[2--10]{keV} X-ray domain). However, this model, put forward well before the Fermi satellite was launched, underestimates by far the measured luminosity at MeV and GeV energies from 2FGL~J1745.6-2858, as the typical spectral shape of IC emission exhibits a pronounced peak-like structure (rather than a powerlaw shape that extends over four orders of magnitude in energy, cf.\ Fig.~\ref{Fig:GCCombinedSpectrum}). \cite{Chernyakova:2011} therefore argue that, if both \hgc\ and 2FGL~J1745.6-2858 are driven by the same emission mechanism, a PWN scenario is likely excluded.

\subsubsection{Gamma-ray emission scenarios involving Sgr~A*}
\label{sec:SgrAEmission}
Its low bolometric luminosity renders Sgr~A* an
unusually quiet representative of galactic nuclei. At the same time,
this property makes the immediate vicinity of the supermassive BH transparent to
high-energy $\gamma$-rays: due to the lack of dense IR radiation fields which would absorb high-energy $\gamma$-rays by pair production processes, photons with energies of up to several TeV are expected to escape a region as small as several Schwarzschild radii in diameter around the BH almost unabsorbed \cite{Aharonian:2005ti}, making HE and VHE $\gamma$-rays a unique probe for particle acceleration and radiation processes close to the BH event horizon.

A multitude of processes is suggested by which Sgr~A* (either directly or indirectly) might produce populations of relativistic particles, leading to $\gamma$-rays of energies up to several ten TeV and possibly beyond. Models can be roughly categorised by the type of particles accelerated (electrons or protons), the physics of the acceleration process, and finally the process of $\gamma$-ray production (by an interaction of the accelerated particles with the ambient magnetic
field or matter). From an observer's standpoint, experimental data can be confronted with model predictions such as the proposed energy flux distribution of the photons, their spatial morphology and predicted temporal variability. Given the modest angular resolution and limited flare sensitivity of current $\gamma$-ray instruments, properties like spectral shape and total energy output currently provide the best way to rule out some models or fine-tune the parameters of others. Scenarios which do not contradict findings at $\gamma$-ray energies and emission at longer (radio, IR and X-ray) wavelengths include 
\begin{itemize}
\item Models in which the $\gamma$-ray production happens close to the BH event horizon. Possible scenarios might include \cite{Aharonian:2005ti} hadronic production processes like interactions of ultrarelativistic protons accelerated by strong ($\geq \unit[10]{G}$) magnetic fields in the vicinity of the BH with the the IR radiation field or surrounding plasma. As these processes are not efficient in transferring proton energy to $\gamma$-rays, an acceleration power of up to $\unit[10^{39}]{erg/s}$ might be required. 
Leptonic processes which transfer the energy of relativistic electrons to $\gamma$-rays by IC and curvature radiation are more efficient, but need well-ordered magnetic or electric fields to prevent radiation losses during the electron acceleration. All models predict correlated X-ray/$\gamma$-ray variability (though at different relative strength). A leptonic scenario with two different electron populations \cite{Chernyakova:2011} could in principle account for both HE and VHE emission, while the other models can explain the VHE emission, but underpredict the measured flux of 2FGL~J1745.6-2858.

\item Models which predict the $\gamma$-ray production to happen within a $\sim\unit[10]{pc}$ zone around the BH due to the interaction of run-away protons with the ambient medium \cite{Aharonian:2005b,Liu2006a,Wang:2009,Ballentyne:2011,Chernyakova:2011,Fatuzzo:2012,Linden:2012a}. The protons are either injected continuously into the medium or impulsively during an event of sudden increase in accretion onto \astar, e.g.\ due to the disruption of a nearby star by tidal forces. Due to the relatively large time (compared to typical flare durations) that the particles need to diffuse out of the central BH region, these scenarios effectively decouple both the time scales and intensity of X-ray flares from those at $\gamma$-ray energies, explaining naturally the absence of significant variability in the HE and VHE data. Some of these models explain the complicated transition from the HE to the VHE part of the $\gamma$-ray spectrum (Fig.~\ref{Fig:GCCombinedSpectrum}) by energy-dependent diffusion time scales and/or different scattering zones where the $pp$ interactions take place.

\item Models in which electrons are either accelerated in termination shocks driven by strong winds emerging from within a couple of Schwarzschild
radii \cite{Atoyan2004} or are injected into the surroundings of \astar, where they produce the observed infrared and X-ray flares by synchrotron radiation, and from which they eventually escape \cite{Kusunose:2012}. These models predict the production of TeV and MeV/GeV $\gamma$-rays, respectively, by IC scattering. Because of the bump-like spectral shape of the $\gamma$-rays produced in this process, these scenarios cannot account for the HE and VHE emission at the same time (see also the discussion in the context of the PWN G359.95-0.04 in section~\ref{sec:PWN}).

\item A hybrid model \cite{Guo:2013}, in which a mix of relativistic protons and electrons are accelerated during a sudden event of increased accretion. By diffusion processes, the protons and electrons move away from the vicinity of the BH and interact with surrounding gas or radiation fields, respectively, thereby explaining the observed $\gamma$-ray emission across the entire energy range.
\end{itemize}

While some of these models suggest correlated
multi-wavelength variability, others predict a steady $\gamma$-ray flux because of diffusion of accelerated particles into the surroundings or acceleration in an extended region far away from the BH event horizon. Therefore, non-observation of variability does not rule out Sgr~A* as a counterpart candidate in general. On the other hand would the detection of variability in the $\gamma$-ray data immediately point to $\gamma$-ray production in the BH vicinity. The most convincing signature would be the discovery of correlated flaring in X-rays (or near IR) and VHE/HE $\gamma$-rays. Such searches have been carried out. In a coordinated multi-wavelength campaign \cite{Aharonian:2008a}, the region enclosing \astar\ was observed both by the Chandra satellite and the H.E.S.S.\ instrument. During this campaign, a major X-ray outburst was detected from the direction of \astar, and a factor 9 increase in the X-ray flux was measured. The total X-ray flare lasted for \unit[1600]{s}; during this time interval, the VHE $\gamma$-ray flux stayed constant within errors, excluding a doubling of the VHE flux at 99\% CL. When interpreting the X-ray flare as synchrotron emission of TeV electrons, the non-observation of an IC VHE counterpart implies a magnetic field strength in the emission region of $\unit[>50]{mG}$ \cite{Aharonian:2008a}. Similar long-term searches have been carried out by \cite{VERITAS:2014}. So far no strong constraints can be derived from these findings, because even larger magnetic fields are expected to be present close to Sgr~A*.

\subsubsection{Cumulative $\gamma$-ray emission from the inner \unit[10]{pc} region}
Due to the fact that the point spread function of $\gamma$-ray instruments covers a region of several \unit{pc} diameter at the GC distance, the observed $\gamma$-ray emission could in principle stem from a combination of multiple emitters (like e.g.\ the discrete objects presented above). In particular, as was argued recently \cite{Bednarek:2013}, the emission might be explained by the cumulative effort of a few thousand millisecond pulsars (MSPs) which could be present in the central star cluster surrounding \astar. While it is not clear whether or not such a population of MSPs is present within the central \unit{pc} of the GC, globular clusters contain a large number of these objects and are established GeV $\gamma$-ray emitters \cite{Abdo:2010b}. In the model proposed by \cite{Bednarek:2013}, the MSP population at the GC could be the result of a past merger of globular clusters, and the HE and VHE emission due to the MSPs themselves and inverse-Compton scattered electrons accelerated in the wind regions of the MSPs. As the total emission is resulting from a large number of different sources spread over a \unit{pc} size region, no variability is expected in this scenario.

\subsection{The Central Molecular Zone in $\gamma$-ray light}
\begin{figure*}
  \includegraphics[width=\textwidth]{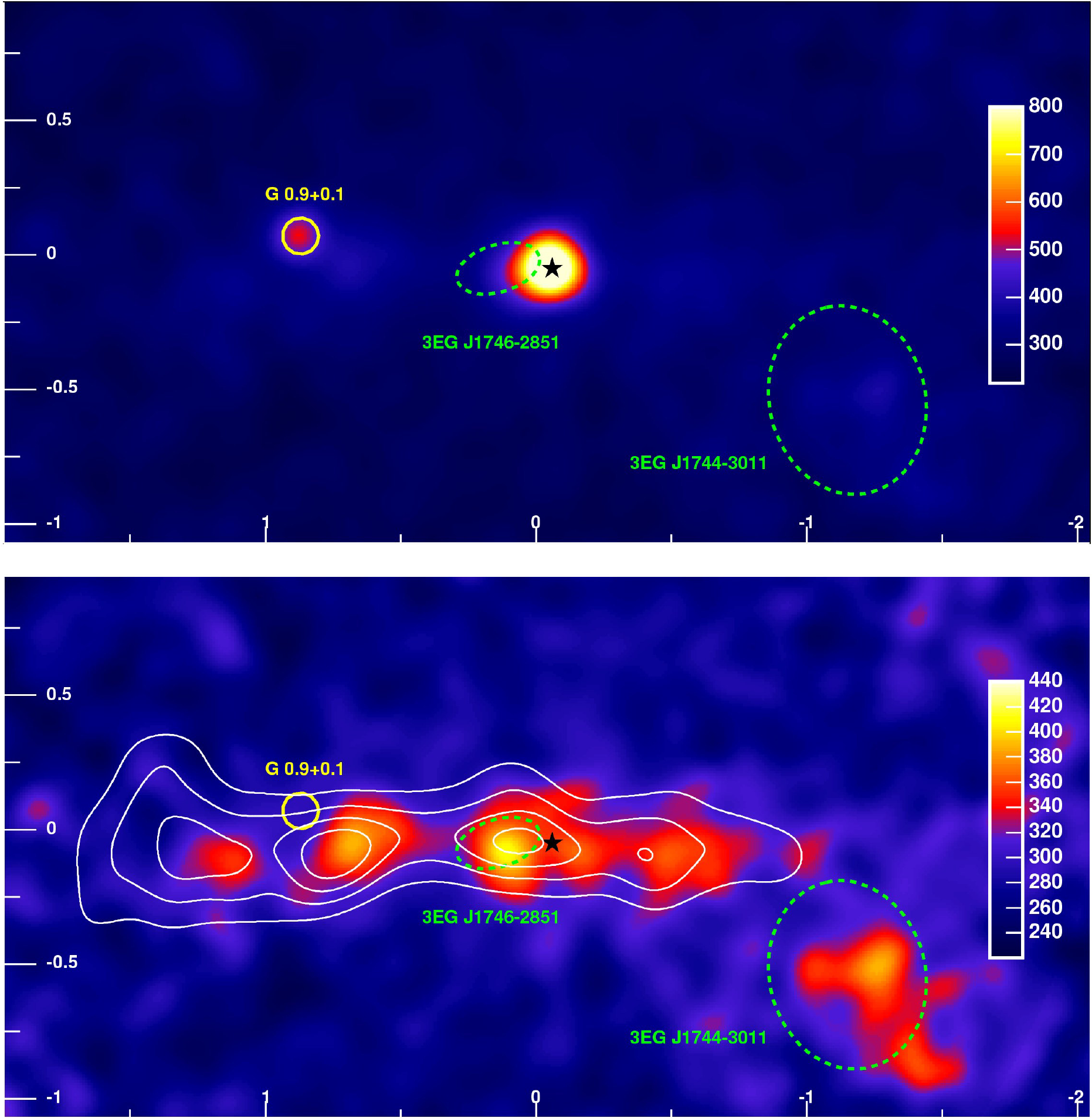}
  \caption{\hess\ VHE $\gamma$-ray images of the Galactic Centre region
    \cite{Aharonian:2006au}. Top: smoothed $\gamma$-ray count map (above a $\gamma$-ray energy of \unit[380]{GeV}, without
    background subtraction) showing emission from the GC point source \hgc\ 
    and the SNR/PWN G0.9+0.1. The position of \astar\ is marked by the black star. Bottom: same map after
    subtraction of the two (assumed point-like) sources, showing an
    extended band of highly significant (14.6 standard deviations above the background) \gr\
    emission and the unidentified source HESS~J1745-303 (south-west of \astar). The white  
    contours show velocity integrated CS line emission \cite{Tsuboi:1999}, smoothed to
    match the angular resolution of the $\gamma$-ray measurement. The positions of two
    unidentified EGRET sources are shown by green ellipses.} 
  \label{fig:GCDiffuse}
\end{figure*} 

Whilst the early detections of the GC at VHE energies were based on data sets of
limited statistics and/or high energy threshold, and concentrated on studying the properties of the GC point source, follow-up observations
provide a much more detailed picture of the central region, and especially of the CMZ.
Since the linear size of the CMZ translates into a solid angle diameter of $\sim 3^\circ$ on the sky, the FOV of current IACTs is sufficiently large to observe the CMZ with a single (or at maximum a few) different pointings of the instrument, enabling the generation of VHE $\gamma$-ray sky maps of the region. Both the VERITAS and H.E.S.S.\ collaborations have published such maps, the most detailed map being based on \unit[49]{h} of GC observations conducted with the H.E.S.S.\ array in the year 2004 \cite{Aharonian:2006au}.

Fig.~\ref{fig:GCDiffuse} (top) displays the resulting (smoothed) $\gamma$-ray count map above an energy of \unit[380]{GeV}. It shows, besides the already discussed GC point source \hgc, another discrete source $\sim 1^\circ$ to the east of \astar, positionally coincident with the SNR/PWN G0.9+0.1 \cite{Helfand:1987}. Subtracting these two sources (assuming a point-like morphology), a band of diffuse $\gamma$-ray emission along the Galactic Plane gets visible, extending across the entire CMZ (Fig.~\ref{fig:GCDiffuse}, bottom). Besides that, another region of extended $\gamma$-ray emission, denoted HESS~J1745-303 \cite{Aharonian:2006zz}, e\-mer\-ges about $1.4^\circ$ south-west of \astar. Before discussing the diffuse $\gamma$-ray emission in detail, we summarise the current knowledge about the nature of the more compact emission regions. 

\subsubsection{The nature of HESS~J1745-303}
Despite detailed multi-wavelength studies \cite{Aharonian:2008aa,Bamba:2009} which followed the discovery of the $\gamma$-ray source, 
HESS~J1745-303 still belongs to a rather long list of unidentified Galactic VHE $\gamma$-ray sources, for
which the lack of solid counterparts at other wavelengths or source confusion makes a firm identification difficult so far. The rather extended and possibly substructured VHE $\gamma$-ray morphology suggests that the emission might be produced by several objects, among them the expanding shell of the SNR G359.1-0.5 (easily recognised in the radio image, see Fig.~\ref{fig:LaRosa}) which is in contact with dense molecular clouds. Southern parts of the VHE emission are considered being due to the nebulae of one or two energetic pulsars \cite{Aharonian:2008aa}. Furthermore, the emission is at least partially overlapping the EGRET source 3EG~J1744-3011. However, the reported flux level of the latter \cite{Hartmann:1999} is too large to smoothly connect to the VHE flux. Besides that, a hint for flux variability of the EGRET source \cite{Torres:2001} makes it particularly hard to consider a common origin of the HE and VHE emission. However, an analysis of Fermi-LAT data of this region \cite{Chernyakova:2011,Hui:2011} indicates that the HE emission is in fact steady, and that the HE spectrum connects well to that of the northern part of the VHE source. A firm association between the HE and the VHE $\gamma$-ray sources is hampered by the fact that the angular resolution is poor and no clear alignment of the HE emission with one of the VHE source components is observed.

\subsubsection{G0.9+0.1: a $\gamma$-ray Pulsar Wind Nebula}
G0.9+0.1 is a well-known composite SNR with a clear shell-like radio
morphology (see the radio image, Fig.~\ref{fig:LaRosa}). The radio shell has a diameter of 8' and is filled with a bright compact core. Given the angular extension of the shell and assuming that G0.9+0.1 is located at the GC at $\sim\unit[8]{kpc}$ distance, the supernova event took
place a few thousand years ago. This notion is supported by the recent discovery of a highly energetic pulsar (PSR~J1747-2809) in the centre of the SNR \cite{Camilo:2009}, with an estimated age of $\sim\unit[2-3]{kyr}$. Based on its dispersion measure, however, the pulsar is possibly located well behind the GC (at a distance of  $\sim\unit[13]{kpc}$, although a location at the GC cannot be excluded). X-ray studies of G0.9+0.1 \cite{Gaensler2001, Porquet2003a} have identified the radio core as a pulsar wind nebula. Spectral softening with increasing distance to the core's centre is found, suggesting the presence of a population of relativistic electrons which cool due to synchrotron radiation on their way outwards. Only faint X-ray emission has been detected from the SNR shell \cite{Porquet2003a}, and no MeV/GeV emission has been found so far.

The morphology of the VHE emission is 
compatible with being a point source centred on the PWN position determined from the Chanrda data. An upper limit of 1.3' at 95\%~CL is derived on the size of the emission region, excluding particle acceleration in the SNR shell as the main source of the VHE $\gamma$-rays \cite{Aharonian:2005br}, such that an identification of the $\gamma$-ray source with the PWN seems compelling. The $\gamma$-ray spectrum extends up to an energy of \unit[6]{TeV} and is best described by a simple powerlaw with a spectral index of $\Gamma=2.40$ (typical for PWNe and Galactic $\gamma$-ray sources in general). The total power radiated in \unit[0.2--10]{TeV} $\gamma$-rays is
$\unit[2\times 10^{34}]{erg~s^{-1}}$, which is the at date second largest TeV luminosity measured for a PWN. The spectral energy distribution of the PWN, including its $\gamma$-ray spectrum, fits a simple one-zone IC scenario, yielding a reasonable magnetic field strength in the nebula of $\sim\unit[6]{\mu G}$ and an interstellar radiation field density of $\unit[5.7]{eV~cm^{-3}}$ \cite{Aharonian:2005br}, assuming that the PWN is located in the GC. In a recent study \cite{Holler:2012}, a radial multi-zone model is applied, in which the lepton population injected at the core of the PWN is propagated outwards. The predicted synchrotron flux in the various zones is matched to that of spatially resolved X-ray spectra, and the VHE $\gamma$-ray emission is predicted. Adopting a distance of the PWN of \unit[13]{kpc}, the authors conclude that IC emission of relativistic electrons cannot account for the observed VHE emission, as predicted fluxes are too low by at least an order of magnitude.

\subsubsection{Diffuse VHE $\gamma$-ray emission}
\label{sec:Diffuse}
The VHE $\gamma$-ray emission detected by H.E.S.S.\ in the CMZ provides important insights into an understanding of the large-scale high-energy astroparticle physics processes at work in the central few \unit[100]{pc} of the Milky Way. The observed emission (Fig.~\ref{fig:GCDiffuse}, bottom) spans a region of roughly $2^\circ$ in galactic longitude with an rms width of about $0.2^\circ$
in galactic latitude. Assuming that the emission is produced in the GC region, the angular extension translates into a projected physical size of about
$\unit[300\times 30]{pc^2}$. When interpreting the emission to learn about the underlying particle population, its propagation through the medium and the acceleration mechanisms at work, the following properties extracted from the VHE observations \cite{Aharonian:2006au} must be taken into account:
\begin{itemize}
\item The spatial extent of the $\gamma$-ray emission is very similar to that of the giant molecular clouds present in the CMZ (cf.\ section \ref{sec:stage}). Indeed, as can be seen from Fig.~\ref{fig:GCDiffuse}, at least for $|l|\leq 1^\circ$ there is a strong correlation between the morphology of the observed
VHE \grs\ and the density of molecular clouds (traced by CS emission). This fact could point to a scenario where locally accelerated cosmic rays of up to several \unit[10]{TeV} energy interact with the material in the clouds, giving rise to the observed emission via $\pi^0\to\gamma\gamma$ decays. Despite the presence of molecular material, no significant emission is however seen in the very eastern part of the CMZ ($|l|\gtrsim 1^\circ$), which may eventually put constraints on the time scale that the high-energy particles had to propagate into the region.
\item It seems unlikely that a distribution of electron accelerators (such as PWNe), clustering similar to the gas density, can account for the observed emission. Given that the large-scale magnetic field strength in the region probably exceeds $\unit[50]{\mu G}$ \cite{Crocker:2010}, electrons of several TeV energy would rapidly cool via synchrotron radiation, such that their VHE $\gamma$-ray emission would appear more compact in the map than is observed. 
\item The reconstructed $\gamma$-ray spectrum
integrated within the longitude-latitude window $|l|\leq 0.8$, $|b|\leq 0.3$ is well-described by
a (hard) powerlaw with spectral index $\Gamma=2.29$,
in agreement with the index observed for the GC source \hgc. 
\item The energy
necessary to fill the entire region with cosmic rays can be estimated
from the measured VHE $\gamma$-ray flux and amounts to $\sim\unit[10^{50}]{erg}$ above an energy of \unit[1]{GeV}, close to the energy transferred into cosmic rays in a typical galactic supernova event. Hence, it seems plausible that the total energy could be provided by a single (and local) accelerator of cosmic rays.
\end{itemize}

Given the similarity between the spectral indexes of the $\gamma$-ray spectra of \hgc\ and the surrounding diffuse emission, it is tempting to explain the diffuse $\gamma$-ray morphology by the interaction of high-energetic protons stemming from the same acceleration process that is ultimately responsible for the emission of \hgc\ (see section\ \ref{sec:HGC} for a discussion of possible counterparts of \hgc). Protons might then escape from the acceleration region and penetrate the surrounding medium, giving rise to the diffuse emission. This is the standard scenario brought up by the H.E.S.S.\ collaboration \cite{Aharonian:2006au}, and has been investigated (and challenged) by a large number of different studies, mostly in the context of diffusion processes in the central \unit[10]{pc} region and within the turbulent CMZ magnetic field. However, before putting the diffuse VHE emission into an astrophysical context, high-energy $\gamma$-ray emission on even larger scales has to be discussed.

\subsection{MeV and GeV diffuse $\gamma$-ray emission}
\begin{figure}
\includegraphics[width=0.48\textwidth]{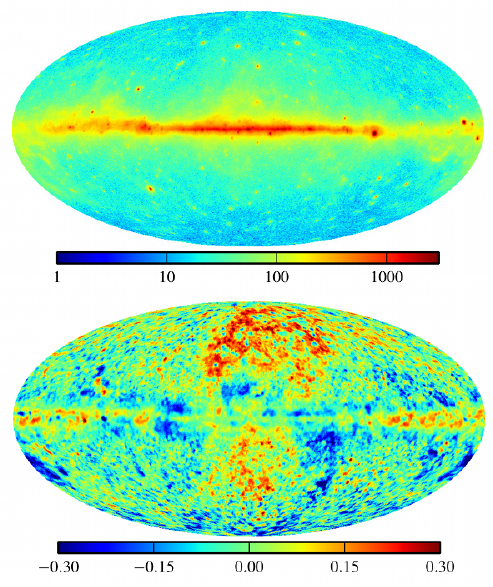}
\caption{Top: $\gamma$-ray count map (energy range \unit[200]{MeV}-\unit[100]{GeV}) obtained from 21 months of Fermi-LAT observations in Galactic coordinates. Diffuse emission is dominating in the Galactic Plane. Bottom: Fractional residual map, i.e.\ (data-model)/data, for the same energy range. The Galactic Plane emission at large is well accounted for by the diffuse model, with some localised regions where the model overpredicts or undershoots observations by up to 30\%. Additionally, there are two large scale regions above and below the Galactic Plane where emission is found which is not predicted by the model. This emission is, for the most part, associated with bipolar outflows from the GC (the so-called the \emph{Fermi bubbles}). Plots taken from \cite{Ackermann:2012}.}
\label{fig:FermiDiffuse}
\end{figure}

At HE energies, the Galactic $\gamma$-ray sky is dominated by diffuse emission produced by relativistic (cosmic ray) particles in interactions with ambient material or radiation fields. The EGRET mission already provided a detailed view of the $\gamma$-ray emission from these processes (for a review on EGRET results, see e.g.\ \cite{Thompson:2008}). With the successful launch of the Fermi satellite, observations of diffuse emission have reached a new level of quality in terms of sensitivity, energy coverage and instrument resolution. To extract information about the cosmic ray population and the properties of the interstellar medium, models are fit to the $\gamma$-ray data which describe the injection of relativistic particles into and their transport through the interstellar medium, as well as the $\gamma$-ray emission they produce. A widely used propagation code is GALPROP (see e.g.\ \cite{Moskalenko:2012} and references therein). Recently, the Fermi-LAT Collaboration presented a detailed comparison of LAT data with GALPROP predictions, showing that the model is able to describe the $\gamma$-ray emission-at-large in most regions of the sky satisfactorily \cite{Ackermann:2012}. Fig.~\ref{fig:FermiDiffuse} shows the all-sky $\gamma$-ray count map obtained by the LAT instrument in 21 month of operation, as well as a residual map (with known $\gamma$-ray sources and a particular realisation of diffuse emission model subtracted), strengthening confidence that the HE diffuse emission of the Galaxy-at-large is reasonably understood. However, this does not completely hold for  inner part of the Galactic Plane, notably the GC region, where the model significantly underestimates the observed $\gamma$-ray flux \cite{Ackermann:2012}. In particular in the GC region, modelling of diffuse emission is facing large systematic uncertainties, because particle populations, interstellar medium and accelerating sources are, if at all, only poorly understood.

Despite these uncertainties, diffuse emission models can be used to probe the $\gamma$-ray sky on large scales, as any residual (after model subtraction) emission potentially points to a yet undiscovered source of $\gamma$-rays. Substantial attention was received in recent years when studies, by subtracting known sources of $\gamma$-ray emission from the Fermi-LAT all-sky maps, revealed the existence of large ($\sim 50^\circ\times 40^\circ$ in Galactic latitude and longitude, respectively) lobes of $\gamma$-ray emission towards the north and south of the GC (the \emph{Fermi bubbles}, visible in the bottom part of Fig.~\ref{fig:FermiDiffuse}), oriented perpendicular to the Galactic Plane \cite{Dobler:2010,Su:2010}. The $\gamma$-ray flux within the bubbles is rather homogeneously distributed, with sharp edges separating the bubbles from the interstellar medium \cite{Su:2010}, and in the energy range of \unit[1--100]{GeV} a hard powerlaw spectrum with spectral index $\Gamma\sim 2$ is found, cutting off at lower and higher energies, with no significant spatial variation across the emission region \cite{Su:2010,Dobler:2010,FermiBubbles:2014}. The bubbles align well with previously identified large-scale emission at microwave energies (the \emph{WMAP haze}, \cite{Finkbeiner:2004,Dobler:2008}, recently confirmed by Planck measurements \cite{Planck:2013h}). Furthermore, the edges of the $\gamma$-ray bubbles coincide with $\sim\unit[2]{keV}$ X-ray emission features present in ROSAT survey maps \cite{Snowden:1997}, and ridge-like structures of polarised radio emission are observed across the bubbles' interiors \cite{Carretti:2013}. 

As has been shown by \cite{Su:2010,Dobler:2010,FermiBubbles:2014}, IC and synchrotron emission from relativistic electrons can simultaneously account for the WMAP haze and the Fermi bubble emission, but the authors also note that the electron population itself must then feature a hard spectrum, disfavouring steady-state Galactic disk electrons which already suffered spectral steepening during their propagation in the Galaxy. Alternatively, the $\gamma$-ray emission could stem from relativistic protons. \cite{Su:2010}, however, argue that the emission would be expected to trace the (presumably not uniform) density of gas in the bubbles' interiors, and that it is unlikely that the emission would then feature a homogeneous surface brightness as observed.
The orientation and location of the bubbles and the fact that the emission is seen all the way down to the inner Galactic Plane suggests that the relativistic particle population is connected to astrophysical processes taking place in vicinity of the GC (implying that the structures extend more than \unit[10]{kpc} above and below the plane). The total $\gamma$-ray luminosity of the bubbles is $\sim \unit[4\times 10^{37}]{erg~s^{-1}}$, and a rough estimate (with significant uncertainties due to the unknown expansion rate of the bubbles and their gas density) of the total energy content and age of the bubbles yields $\sim\unit[10^{54}-10^{55}]{erg}$ and $\sim \unit[10^7]{yr}$, respectively \cite{Su:2010}. Cocoon and jet-like features have been claimed present in the Fermi-LAT data \cite{Su:2012} (but see the recent study by \cite{FermiBubbles:2014}). These structures, if confirmed, are located within the Fermi bubbles and, like the bubbles, exhibit a hard spectrum ($\Gamma\sim 2$) in the \unit[1-100]{GeV} energy range, possibly suggesting that the Fermi bubbles have their origin in the existence of the jet.

\subsection{Towards a global picture of the Galactic Centre high-energy astrophysics}
The recent results obtained from HE and VHE $\gamma$-ray observations of the GC region and beyond can be used to set up and constrain models which try to explain the emission in a global astrophysical context. Here, the activity of the central BH \astar\ plays a dominant role, but also do (former) periods of star formation within the central few \unit[100]{pc} of the Galaxy. Ideally, such modelling would address (i) the total energetics of the region, (ii) the nature of the $\gamma$-ray emission of \hgc/2FGL~J1745.6-2858, (iii) the origin of the diffuse VHE $\gamma$-ray emission and its correlation with molecular material, and (iv) the origin of the emission seen from the Fermi bubbles. It is clear that such model is challenging given the complexity of the GC region, and we are far from understanding all the ingredients in this complicated interplay of the different constituents and processes at work.

\subsubsection{On the origin of the Fermi bubbles}
Various scenarios have been put forward to explain the origin of the Fermi bubbles, connecting the feature with past activity of the central few $\unit[100]{pc}$ region. These include
\begin{itemize}
\item An epoch of enhanced accretion onto the BH \cite{Su:2010}, producing AGN-like jet emission \cite{Guo:2012}. Even at a moderate mass accretion rate of $\sim 1\%$ of the BH's Eddington luminosity, an accretion event lasting for $\sim\unit[10^5-10^6]{years}$ would account for the total energy stored in the bubbles.
\item Star formation activity and corresponding supernova explosions driving an energetic wind, which is considered powerful enough to provide the required energy \cite{Su:2010,Crocker:2011a,Carretti:2013}.
\item Periodic star accretion onto the BH, driving hot plasma outflows \cite{Cheng:2011}. 
\end{itemize}

Much harder to address than the energetics are the morphology and spectral properties of the bubble emission: assuming that the emission is produced by a population of relativistic electrons with a hard energy spectrum, a mechanism must be in place to transport the electrons from their acceleration region (close to the GC) into the bubbles. Such transport mechanism must act on smaller time scales than the typical cooling time of $ \unit{TeV}$ electrons ($\sim\unit[10^5-10^6]{years}$) to avoid significant steepening of the electron spectrum. This excludes classical diffusion processes as a mechanism for electron transport. However, advection of relativistic electrons into the bubble regions by powerful jets emanating from a past activity of the BH, or in-situ acceleration of electrons by internal shocks within the jets appear to be viable alternatives \cite{Su:2010,Yang:2012}. \cite{Yang:2012} use 3D hydrodynamic simulations of bipolar jet outflows to show that due to projection effects the physical size of the bubbles might be smaller than first suggested ($\sim\unit[6]{kpc}$ rather than \unit[10]{kpc}), possibly allowing electrons to be fully advected into the bubbles before significant cooling sets in. Besides that, the predicted IC $\gamma$-ray morphology of the simulated bubbles is well in agreement with the observed homogeneous surface brightness. 

Star forming events in the CMZ can possibly explain the polarised radio emission detected from the bubbles' interior \cite{Carretti:2013}. Given the lobes' microwave, radio, X-ray and $\gamma$-ray emission, it is proposed that the bubbles host a population of relativistic electrons, first accelerated by SN events in the CMZ, and subsequently advected into the bubbles by winds with a speed of up to $\unit[1000]{km/s}$.

Finally, \cite{Cheng:2011} propose that periodic star capture and disruption injects hot plasma into the Galactic halo, with an estimated total power of $\sim \unit[3\times 10^{40}]{erg~s^{-1}}$. Acceleration of electrons would then be possible by internal shocks in the plasma, and the homogeneous morphology as well as the energy spectrum of the $\gamma$-ray emission is achieved by a superposition of many repeated star disruptions. This scenario is questioned by \cite{Mertsch:2011} who instead propose that the electrons are accelerated in-situ in the bubbles by a 2nd order Fermi acceleration process.

Besides electrons, relativistic protons are put forward to explain the bubbles' $\gamma$-ray emission. As protons do not suffer from cooling, there are no strong constraints on the time scale of their propagation from the acceleration region into the bubbles' interiors, and time scales of $\sim\unit[10^6-10^9]{years}$ are proposed. While \cite{Su:2010} argue that the flux of $\gamma$-ray emission from proton-proton interactions should scale with the gas density in the bubbles, and is therefore unlikely to explain the homogeneous surface brightness, \cite{Crocker:2011a} suggest that the protons stay confined to the bubble, in which case the $\gamma$-ray luminosity due to proton-gas collisions is independent of the gas density. Cosmic ray protons could be injected into the bubbles by AGN jet emission as e.g.\ proposed by \cite{Guo:2012}. At first sight, a jet emission scenario does not explain in a straightforward way the rather extended (in Galactic longitude) morphology of the bubbles, as for jet emission a more cocoon-like shape would be expected. However, the simulations by \cite{Guo:2012}, in which bipolar jets are injected perpendicular to the Galactic plane, suggest that the jet structure is expanding in lateral direction due to overpressure in the jet plasma compared to the interstellar medium, and that the observed morphology of the bubbles can be matched rather well. \cite{Zubovas:2011} argue that the Fermi bubbles can be explained by a bright accretion event coincident with an epoch of extensive star formation $\sim \unit[6\times 10^6]{years}$ ago. A quasi-spherical outflow is proposed by these authors, which is compressed laterally by the gas pressure of the Galactic disk, and therefore forms the bubble-like morphology that is observed. Cosmic ray protons may then be accelerated at shock fronts in regions where the plasma collides with the interstellar medium. 

Addressing the star formation scenario, \cite{Crocker:2011a} suggest that relativistic protons accelerated by SN explosions in the CMZ are advected by a strong wind into the bubble regions. Given the SN rate in the CMZ and their typical explosion energy, an average power as large as $\unit[10^{39}]{erg~s^{-1}}$ could be tranferred into cosmic ray acceleration, which matches the energy loss of the total proton population in the bubble (as  extrapolated from the HE surface brightness under the assumption that protons are confined to the bubble region).

\subsubsection{On the sources of the diffuse VHE emission}
What is the nature of the parent particle population responsible for the diffuse VHE $\gamma$-ray emission along the CMZ? Given the strong correlation of its surface brightness with the density of molecular material, it is natural to assume that relativistic protons are penetrating the region which interact inelastically with gas molecules and produce $\gamma$-ray emission mainly by subsequent $\pi^0\rightarrow\gamma\gamma$ decays.
However, cosmic ray protons from the Galaxy-at-large cannot account for the observed emission, since the measured $\gamma$-ray flux is both larger (a factor of 3-9 above a $\gamma$-ray energy of \unit[1]{TeV}) and harder (spectral index $\Gamma\sim 2.3$) than expected when the molecular material was bathed in a sea of cosmic ray protons that propagated through the entire Galactic disk. In the latter case, an index of $\Gamma\sim 2.7$ is expected, since for $\gamma$-ray production by proton-proton collisions the spectral index of the $\gamma$-ray spectrum is identical to that of the underlying proton distribution (which exhibits a spectral index of $\sim 2.7$ measured close to the Earth). These findings strongly suggest the presence of one or more proton accelerators close to the CMZ, i.e.\ close to the region where the $\gamma$-ray emission is produced, such that cosmic ray propagation effects which lead to spectral steepening do not play a significant role. 

In the context of identifying the accelerator, the fact
that no emission is seen farther than $|l|\sim 1^\circ$ (i.e.\ no more than $\sim\unit[140]{pc}$ in projection away from the GC) is particularly interesting. In their discovery paper \cite{Aharonian:2006au}, the \hess\ collaboration came up with the
rather simple, yet convincing explanation that the cosmic ray protons may
have been accelerated by an object close to the GC, like e.g.\ \astar or \aeast. The protons would then diffuse away from the accelerator and at some point fill the entire CMZ, giving rise to the observed VHE $\gamma$-ray emission. If, however, the protons were injected by the accelerator only recently, they may not have had enough time to reach the outer parts of CMZ, which would result in the observed lack of VHE emission in these parts. Adopting a cosmic ray diffusion coefficient of $D\sim \unit[10^{30}]{cm^2~s^{-1}}$, typical for TeV cosmic rays in the Galactic disk, an accelerator of age $\sim\unit[10^4]{years}$ can reproduce the observed $\gamma$-ray morphology, and in particular the lack of emission beyond $1^\circ$ distance from the centre. Other studies adopted this idea of a ''diffusion clock'', with the aim of deriving (energy-dependent) diffusion coefficients applicable for the GC environment \cite{Buesching2007,Dimitrakoudis:2009}.

\cite{Buesching2008} try to explain both the morphology of the diffuse emission and the GC point source \hgc\ with a unified
approach in which cosmic rays are accelerated $\sim\unit[5-10]{kyr}$ ago
in the shock wave of the SNR \aeast, leading to the observed diffuse $\gamma$-ray emission in the CMZ.
At some point the shock wave of the SNR is proposed to have collided with \astar, and particle acceleration near the BH was initiated, 
producing the VHE \gr\ emission of \hgc. Assuming that the diffusion coefficient determined for the CMZ is valid also in the central \unit[10]{pc} region, the authors claim that this last round of particle acceleration can only have happened in the recent past ($\sim\unit[100]{years}$ ago) to be consistent with the point-like morphology of \hgc. 

The question of whether or not this simple proton diffusion picture holds in the turbulent magnetic fields of the CMZ can be addressed in simulations which propagate individual protons through the field by applying the Lorentz force directly to individual particle trajectories. This method is used e.g.\ by \cite{Wommer2008} who consider a Kolmogorov-type turbulence for the magnetic field structure and a total magnetic field strengths of $\unit[10-100]{\mu G}$, which is a realistic choice given that a lower limit of $\unit[50]{\mu G}$ was recently derived on scales of the CMZ \cite{Crocker:2010}. At any point in time the diffusing protons may then undergo inelastic collisions with molecular material (with a probability that depends on the gas densitiy) and hence produce $\gamma$-ray emission. From their studies, the authors conclude that relativistic protons injected at the inner GC cannot diffuse out to distance scales of a degree, since they scatter with the ambient medium (and get lost) on much smaller scales (maybe producing the emission seen from \hgc). Hence, the idea that a central accelerator is responsible for the observed diffuse $\gamma$-ray morphology appears difficult to realise. Using the same arguments, this also holds for a collection of individual particle accelerators distributed along the CMZ, which would produce a $\gamma$-ray map with much more compact emission close to the position of these accelerators. \cite{Wommer2008} conclude that only stochastic intercloud acceleration of a hard-spectrum proton distribution can possibly produce the observed $\gamma$-ray morphology. The feasibility of this idea in terms of the achievable acceleration efficiency was further developed in various studies \cite{Melia:2011,Fatuzzo:2012a,Amano:2011}, with the result that stochastic acceleration should in general allow proton acceleration to multi-TeV energies. 

Still, it remains an open question how the relativistic particles get injected into the intercloud medium in the first place. \cite{Crocker:2011b} show that the VHE diffuse emission can be explained by protons accelerated within the remanants of supernovae taking place in the CMZ. Due to the presence of a strong wind, the propagation of these protons is not entirely diffusion-driven, e.g.\ implying that the relativistic particles cannot penetrate fully into the dense molecular clouds before they get advected away by the wind (and ultimately may contribute to a proton population filling the Fermi bubble regions). More sensitive observations are needed to ultimately clarify which of the above discussed scenarios to explain the VHE \gr\ view of the CMZ is correct.


\section{Part II. Searches for GeV and TeV dark matter signatures in the Galactic Centre region}
\label{sec:part2}

It is a widely accepted idea that dark matter, which contributes about 27\% to today's total energy content of the universe \cite{Planck:2013}, consists of elementary particles. Dark matter has so far been detected only very indirectly due to the impact of its mass on the surroundings: Zwicky's original discovery of dark matter \citep{Zwicky:1933}, for example, is based on the fact that the gravitational field provided by the luminous matter of the Coma cluster is not large enough to keep the individual galaxies bound to the cluster, asking for the presence of an additional (very large) non-luminous mass concentration. Similarly, the large radial velocities of stars within galaxies can be explained consistently by the presence of extended dark matter halos, providing enough mass to keep the stars bound. Today's probably best proof of the existence of dark matter is obtained from observations of the Bullet cluster \cite{Clowe:2006}, from which not only the spatial distribution of dark matter could be reconstructed (using the gravitational lensing effect), but it was also shown that the dark matter is essentially collisionless, i.e.\ interacts (apart from gravitation) at best only very weakly with itself and the surrounding matter. 

Since dark matter particles have not been directly detected so far, they must be electrically neutral, stable (or have a lifetime large compared to the age of the universe) and non-baryonic. Comparison of simulations with data from redshift surveys \citep{Springel:2006} suggest that at least a prominent fraction of the dark matter must be in the form of massive, i.e.\ non-relativistic, particles (\emph{cold dark matter}, CDM). From the expansion history of the universe, a velocity-weighted dark matter self-annihilation cross section at thermal freeze-out of
\begin{equation*}
\langle\sigma v\rangle \sim \unit[3\times 10^{-26}]{cm^3/s}
\end{equation*}
is deduced ($v$ is the relative velocity of the DM particles in the annihilation process). This cross section is of the same order as a typical weak cross section. Taken altogether, this suggests that the bulk of dark matter consists of weakly interacting, massive particles (WIMPs) of mass $m_\mathrm{DM} \gtrsim \unit[10]{GeV}$. An upper bound to the WIMP mass of $m_\mathrm{DM}\lesssim \unit[340]{TeV}$ is suggested by \cite{Griest:1990}, based on partial wave unitarity arguments.

Annihilation or decay of WIMPs into standard model particles can lead, among others, to the production of \grs\ in the energy range of \unit{GeV} up to several \unit{TeV} (depending on the unknown mass of the WIMPs), which can in principle be detected by \gr\ satellites or ground-based \gr\ telescopes, respectively. To maximize the probability for a detection, searches for such a signal are carried out best in sky regions of enhanced DM (energy) density $\rho_\mathrm{DM}$. Ideally, nearby regions are chosen, as the \gr\ flux scales with the distance $s$ to the region as $s^{-2}$. At the same time, the sensitivity of any search is enhanced in regions which exhibit only modest astrophysical \gr\ background. Searches therefore concentrate mainly on exploring the cores of dwarf galaxies, globular clusters and galaxy clusters (almost no astrophysical background, but at large distances of $\sim\unit[10]{kpc}- \unit[100]{Mpc}$), the Galactic centre (close-by, but strong contamination by astrophysical sources) and the Galactic halo (a compromise between possible contamination and a small distance). It should be noted that all searches are most sensitive to DM annihilation rather than decay, since (assuming that annihilation into standard model particles is possible) the annihilation rate is proportional to $\rho_\mathrm{DM}^2$, whereas in the case of DM decay, the DM density enters the decay rate only linearly. We will therefore restrict our discussion to the annihilation case, although in most studies lower limits on the lifetime of DM particles are derived as well.

The form of the \gr\ spectrum from DM annihilations depends on the assumed annihilation physics. For secondary photon production, a rather broad photon energy distribution up to the DM particle mass $m_\mathrm{DM}$ is expected; however, prompt annihilation into a two-photon or $\gamma Z$ final state, or contributions such as final-state radiation may lead to sharp features in the energy distribution. Assuming isotropy, the differential photon flux from a solid angle $\Delta\Omega$ along a particular observation direction $\vec{n}$ is given by
\begin{equation}
\frac{\mathrm{d}\Phi(\Delta\Omega, E_\gamma)}{\mathrm{d}E_\gamma} = \frac{1}{4\pi} \underbrace{\frac{\langle\sigma v\rangle}{2m_\mathrm{DM}^2} \frac{\mathrm{d}N_{\gamma}}{\mathrm{d}E_\gamma}}_{\text{particle physics}} \cdot \underbrace{J(\Delta\Omega, \vec{n})\Delta\Omega}_{\text{astrophysics}},
\label{eq:CrossSection}
\end{equation} 
where $\frac{\mathrm{d}N_{\gamma}}{\mathrm{d}E_\gamma}$ is the average \gr\ spectrum per annihilation process, summed over all possible final states. The astrophysical factor $J$ denotes the squared DM density integrated along the line of sight (l.o.s.) and averaged over the solid angle $\Delta\Omega$ covered by the observations:
\begin{equation*}
J(\Delta\Omega, \vec{n}) = \frac{1}{\Delta\Omega}\int_{\Delta\Omega} d\Omega \int_{\text{l.o.s.}} ds~\rho_\mathrm{DM}^2(\Omega, s),
\end{equation*}
where $s$ is the distance of the line of sight element to the observer. For DM searches towards the central Galactic DM halo, assuming that the DM density is distributed spherically symmetric around the GC, it follows that
\begin{equation*}
\rho_\mathrm{DM}(\Omega, s) \equiv \rho_\mathrm{DM}(R),
\end{equation*}
where
\begin{equation*}
R(s) = \sqrt{R_{\astrosun}^2 + s^2 - 2 R_{\astrosun} s \cos\psi}
\end{equation*}
is the radial distance from the GC at any point along the line of sight (observed under an angle $\psi$ to the GC), and $R_{\astrosun}=\unit[8]{kpc}$ is the distance of the GC to the observer.

\subsection{$\gamma$-rays from WIMP annihilations}
A broad spectrum of theoretically motivated WIMP candidates exist in the literature (for a review, see e.g.\ \cite{Bertone:2005}). The underlying theories all provide a stable (or at least quasi-stable) particle in the mass range suggested for cold dark matter, with predicted annihilation cross sections of the order of $\langle\sigma v\rangle \sim \unit[3\times 10^{-26}]{cm^3/s}$, required to explain the relic DM density in $\Lambda$CDM cosmology ($\Omega_\mathrm{c}h^2 = 0.12$, see e.g.\ the recent determination from Planck data \cite{Planck:2013}). Among others, these are
\begin{itemize}
\item Little Higgs models, which are minimal effective extensions to the Standard Model (SM) to solve the problem of a radiative instability of the SM Higgs boson mass (see \cite{Schmaltz:2005} and references therein). Several variants of these models can provide suitable and stable (through the conservation of dedicated parity quantum numbers) DM candidates (e.g.\ \cite{Birkedal:2004, Freitas:2009}).
\item Models of universal extra dimensions, in which all SM fields propagate into (one or more) compact extra spatial dimensions (for a review, see \cite{Hooper:2007}). Due to conservation of KK-parity, the lightest Kaluza-Klein particle (LKP) is stable. Assuming the LKP to be the Kaluza-Klein excitation $B^{(1)}$ of the photon (a spin-1 particle), predicted masses are in the range of $m_\mathrm{DM} \sim \unit[400-1200]{GeV}$ \cite{Servant:2003}.
\item Models of Supersymmetry, and here in particular the Minimal Supersymmetric Standard Model (MSSM). In the MSSM, each SM fermion/boson state is complemented by a bosonic/fermionic superpartner, respectively, with identical quantum numbers (but spin). Additionally, a second Higgs doublet must be introduced (for a review on SUSY DM, see e.g.\ \cite{Jungman:1996}). The symmetry between SM particles and their SUSY partners is broken at present collider energies such that only the SM particles (with masses small compared to those of the superpartners) can be produced in the laboratory. If so-called R-parity holds, the lightest supersymmetric particle (LSP) cannot decay into SM particles and is therefore stable. In the MSSM, the LSP is the lightest neutralino state $\tilde{\chi}_1^0$, which is a linear combination of the gauginos $\tilde{W}^3$ and $\tilde{B}$, and the higgsinos $\tilde{H}_1^0$ and $\tilde{H}_2^0$. The mass of this Majorana (spin-$\frac{1}{2}$) fermion is in the range of several $\unit[10]{GeV}$ up to several $\unit[10]{TeV}$, depending on the choice of parameters.
\end{itemize}
In this review, we will focus on the detection of \emph{prompt} $\gamma$-rays, i.e.\ that these photons are produced in the final state of the annihilation process, in contrast to \emph{secondary} photons, which are e.g.\ radiated when final state particles like positrons undergo synchrotron cooling or IC scattering. Despite the large variety of proposed WIMP candidates, most prompt $\gamma$-ray signatures of DM annihilation can be calculated in a model-independent way, since for a given pair of SM particles in the final state, the spectrum of $\gamma$-rays is largely determined by the dynamics of the fragmentation process and possible final state radiation. However, the actual WIMP masses, the annihilation cross sections and the branching ratios into the various SM final states do depend on the individual WIMP models. In most cases, publicly available software can be used to predict those (e.g.\ the DarkSUSY package \cite{Gondolo:2004} for calculations within the framework of supersymmetry).

\begin{figure}[htbp]
\includegraphics[width=0.48\textwidth]{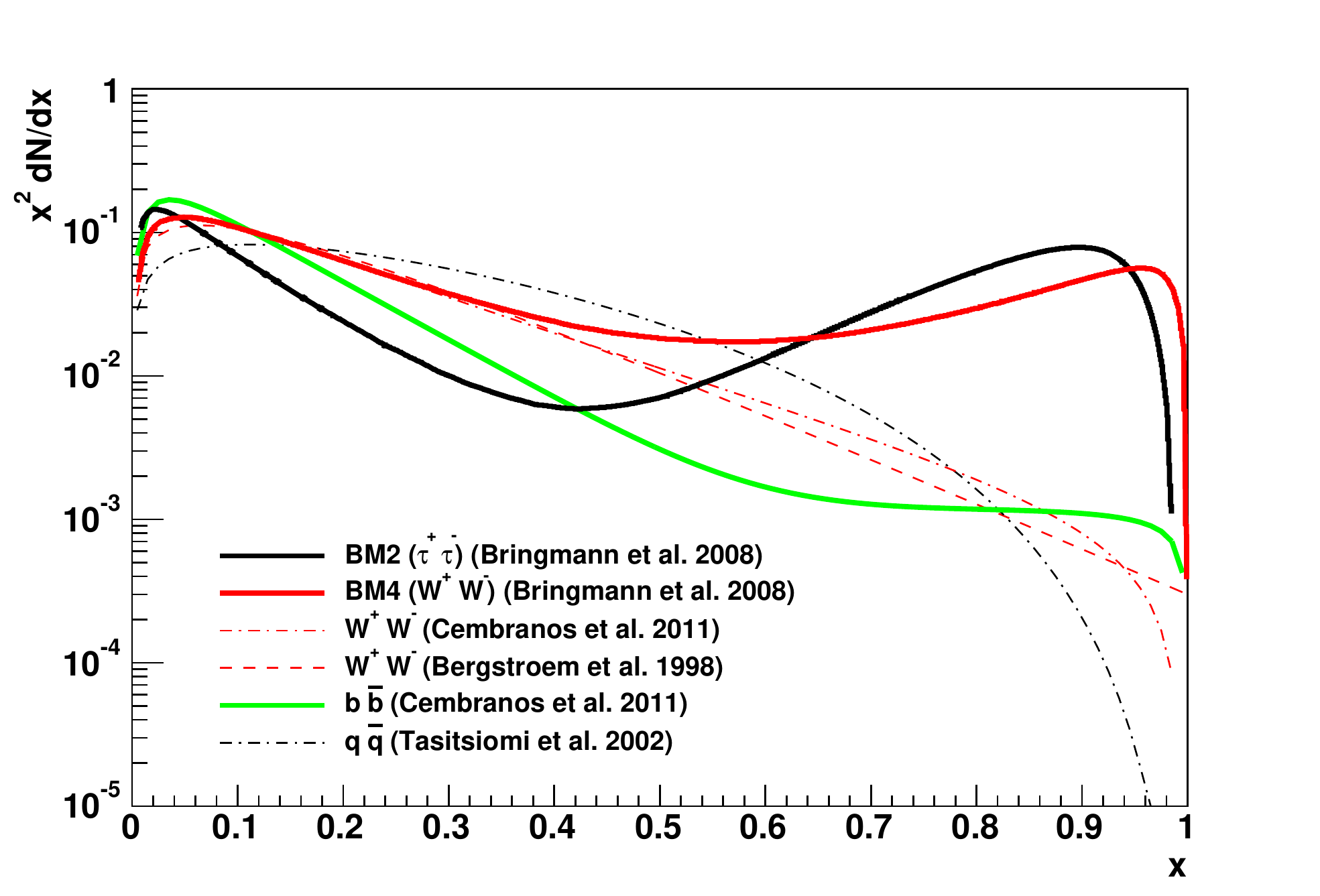}
\caption{Example $\gamma$-ray spectra predicted for neutralino annihilations into $\tau^+\tau^-$, $W^+ W^-$, $b\bar{b}$ and light quark final states. The scaling variable $x$ is defined as $x=E_\gamma/m_\mathrm{DM}$, where $E_\gamma$ is the energy of the photon. The spectrum parametrisation of \citep{Tasitsiomi:2002} does not depend on the DM mass and accounts for photons produced during the hadronisation process only, while the results by \citep{Cembranos:2011} depend on DM mass and take final state radiation into account. Here, $m_\mathrm{DM}=\unit[1]{TeV}$ was assumed. The calculations by \citep{Bringmann:2008} account for internal bremsstrahlung processes which enhance the spectrum towards $x=1$ (i.e.\ for photon energies close to the DM mass). The latter spectra depend strongly on the assumed DM model. Here, the MSSM-based benchmark models BM2 and BM4 of \cite{Bringmann:2008} are shown, with neutralino masses of $m_\mathrm{DM}\sim\unit[800]{GeV}$ and $m_\mathrm{DM}\sim\unit[20]{TeV}$, respectively. For the BM2 neutralino, the main contribution to internal bremsstrahlung comes from the $\tau^+\tau^-$ decay channel, whereas for the BM4 neutralino the $W^+W^-$ channel dominates. Note that the average number of $\gamma$-rays produced in the various final states may differ significantly.}
\label{fig:DMSpectra}
\end{figure}

Prompt $\gamma$-ray emission can be subdivided into three different categories:
\begin{itemize}
\item The aforementioned photon production from final state radiation (FSR) of charged particles in the final state and the fragmentation of final state quarks. The resulting $\gamma$-ray energy spectrum is rather broad, usually peaking well below the DM mass. Fig.~\ref{fig:DMSpectra} shows examples of $\gamma$-ray spectra predicted from the annihilation of neutralinos into final states such as $W^+ W^-$, $\tau^+\tau^-$, $\mu^+\mu^-$ and $b\bar{b}$. A set of recent spectrum parametrisations based on simulations with the PYTHIA package improves on earlier works (e.g.\ \cite{Bergstroem:1998, Tasitsiomi:2002}) in that it characterises in detail the dependence of the spectral shape both on the final state particles and the mass of the DM particle \cite{Cembranos:2011}. Note that the average number of $\gamma$-rays per annihilation event depends strongly on the final state: for the lepton final states of a \unit[1]{TeV} WIMP annihilation, only a few photons per annihilation event are expected due to final state radiation, whereas for quark and gauge boson final states more than 50 $\gamma$-rays are expected from the hadronisation process (for details see \cite{Cembranos:2011}).
\item A direct annihilation into a two particle final state, where at least one of these particles is a photon. Since WIMPs do not couple to photons, the amplitudes of such processes are loop-suppressed, resulting in event rates which are typically smaller by two to four orders of magnitude compared to the first-order process discussed above \cite{Bergstroem:1997, Bern:1997, Bergstroem:2005, Ferrer:2006, Gustafsson:2007}. On the other hand, these processes provide rather distinct features in the $\gamma$-ray energy distribution: for direct annihilation into a two-photon final state, each photon carries an energy $E_\gamma = m_\mathrm{DM}$, leading to a monochromatic line in the $\gamma$-ray spectrum at the WIMP mass. In case of the process $\chi\chi\rightarrow \gamma X$, where $X$ is co-produced particle of mass $m_\mathrm{X}$ (e.g.~the $Z^0$ gauge boson), $E_\gamma = m_\mathrm{DM}\left(1-\frac{m_\mathrm{X}^2}{4m_\mathrm{DM}^2}\right)$. This spectral property is rather unique and easy to distinguish from the usually broad $\gamma$-ray spectra produced by astrophysical (particle accelerating) sources; this is why it is often called a ''smoking gun'' signature.
\item Photons can be radiated from virtual particles inside the annihilation graph. Such radiative corrections to (tree-level) processes are called \emph{internal bremsstrahlung} (IB, \cite{Bergstroem:2005a, Birkedal:2005}). The spectrum of IB photons typically shows a broad, however distinct, edge-like structure close to the mass of the DM particle. The relative $\gamma$-ray flux from this process is suppressed w.r.t.\ the tree-level flux and depends strongly on the DM particle physics model. For some realisations of DM, however, the IB flux is expected to dominate the overall annihilation spectrum close to $E_\gamma=m_\mathrm{DM}$. Examples of $\gamma$-ray spectra for MSSM-based DM models \cite{Bringmann:2008} are shown in Fig.~\ref{fig:DMSpectra}, highlighting the general importance of the IB process to be included in the spectrum calculations.
\end{itemize}

\subsection{Modelling the Galactic Halo}
While the particle physics factor of eq.~(\ref{eq:CrossSection}) can be rather well calculated for most WIMP models, the astrophysical factor $J$ is subject to large uncertainties, since the DM distribution in galaxies and galaxy clusters is only poorly known. This hold especially for the DM halo of the Galaxy: recent N-body simulations of Milky Way-like galaxies, Aquarius \cite{Springel:2008} and Via~Lactea~II \cite{Diemand:2008}, predict a DM distribution which is made up of a smooth halo component (the density of which increases towards the centre), in which a rich spectrum of DM sub-halo clumps of different masses is embedded. Towards the centre of the halo, predicted DM densities are different by factors of a few. At the same time measurements of the density in the inner halo through rotational curves are difficult and therefore lack the needed precision to constrain simulations (see e.g. \cite{Strigari:2013} and references therein). Recent simulations taking into account the feedback of baryonic matter on the halo formation \cite{Pontzen:2012, Governato:2012} suggest that baryonic heating such as stellar feedback or supernova explosions may lead to a flattening of the inner halo DM profile from a cusp to a cored form, and may even shift the peak of the distribution away from the dynamical centre of the galaxy by a significant distance ($\sim\unit[300-400]{pc}$ for a Milky Way-sized galaxy \cite{Kuhlen:2013}).

\begin{figure}[htbp]
\includegraphics[width=0.48\textwidth]{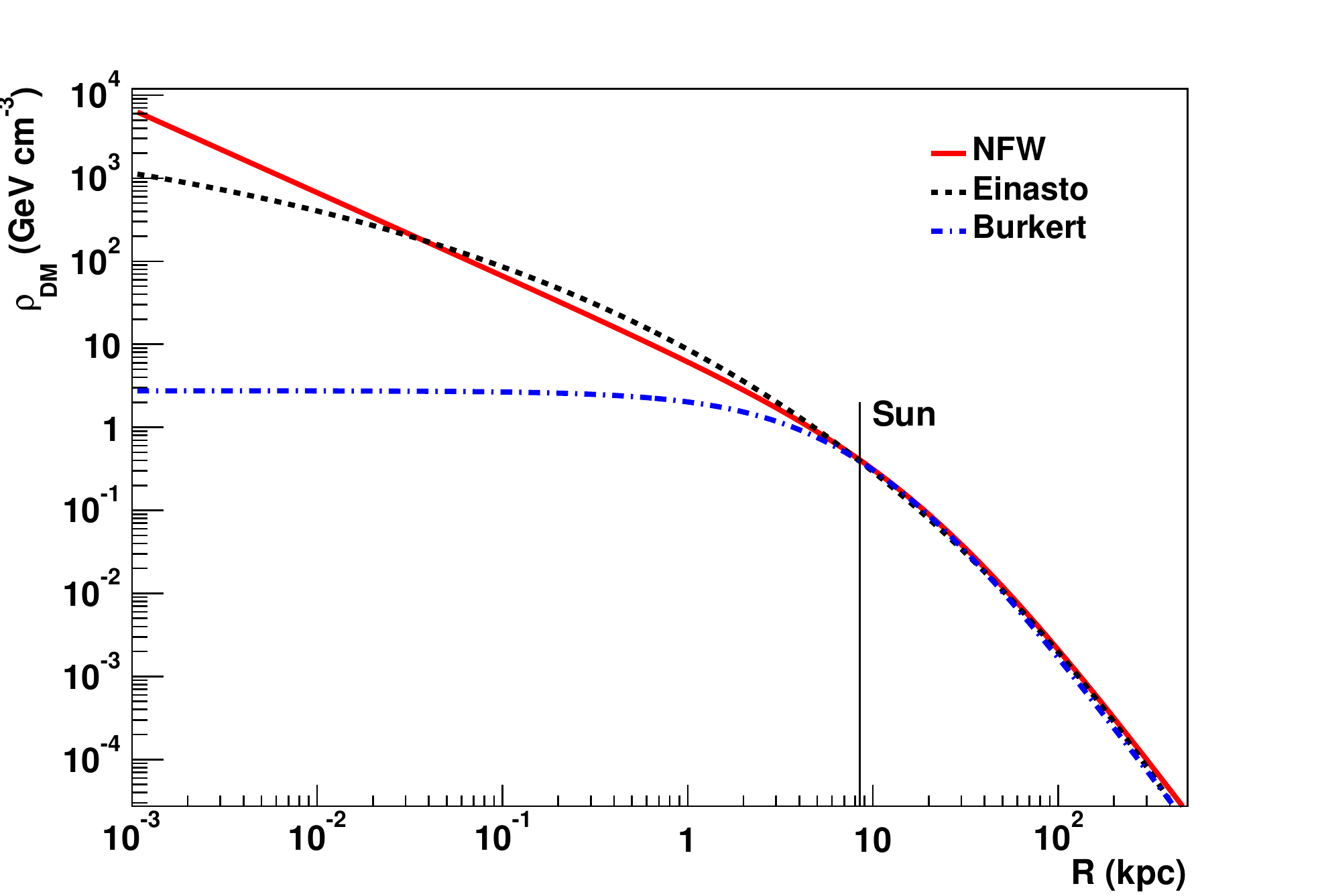}
\caption{Comparison of the profile shape of different proposed DM halo densities of the Milky Way. All curves are normalised to a DM density of $\rho^{\astrosun}_{DM} = \unit[0.39]{GeV/cm^3}$ \cite{Catena:2010} at the position of the Sun (at a distance of $R=\unit[8.5]{kpc}$, denoted by the vertical line). While the NFW and Einasto profiles (full and dashed curves, respectively) predict a cusped DM density, as suggested by N-body simulations, the Burkert profile (dashed-dotted curve) is obtained from a fit to observational data and exhibits a cored shape with constant density towards the GC.}
\label{fig:DMProfiles}
\end{figure}

In the Milky Way, the DM energy density at the very centre may well exceed values of $\unit[10^3]{GeV/cm^3}$, compared to $\rho^{\astrosun}_\mathrm{DM} = \rho_\mathrm{DM}(R=\unit[8]{kpc})=\unit[0.39]{GeV/cm^3}$ \cite{Catena:2010} at the position of the Sun. Leaving aside for the moment complications due to the backreaction of baryonic matter on the DM distribution, the density profiles of the DM-only simulations Aquarius and Via~Lactea~II, assuming radial symmetry, can be parametrised rather well by the universal Einasto and Navarro-Frenk-White (NFW) profiles,
\begin{align}
\rho_\mathrm{DM}^\mathrm{Ein}(R) &= \rho_\mathrm{s}\exp\left\lbrace-\frac{2}{\alpha}\left[\left(\frac{R}{r_\mathrm{s}}\right)^\alpha -1 \right]\right\rbrace\\
\rho_\mathrm{DM}^\mathrm{NFW}(R) &= \rho_\mathrm{s} \left[\frac{R}{r_\mathrm{s}}\left(1+\frac{R}{r_\mathrm{s}}\right)^2\right]^{-1},
\label{eq:Profiles}
\end{align}
respecively, where $R$ is the distance to the GC, and $\alpha$ and the scaling radii $r_\mathrm{s}$ are free parameters. For the Einasto profile, $\alpha=0.17$ and $r_\mathrm{s}=\unit[21]{kpc}$ provide a good fit to the Via~Lactea~II simulation, whereas an NFW profile with $r_\mathrm{s}=\unit[20]{kpc}$ matches the Aquarius simulations \cite{Pieri:2011}. These parametrisations exhibit a steeply increasing DM density towards the GC and are therefore referred to as \emph{cusp profiles}. On the other hand, observations of other spiral galaxies \cite{Gentile:2004} and recent studies of stellar velocities within the Milky Way \cite{Nesti:2013} suggest a rather flat (\emph{cored}) DM profile in the inner part, which can be described empirically by a Burkert profile \cite{Burkert:1995}:
\begin{equation*}
\rho_\mathrm{DM}^\mathrm{Bur}(R) = \rho_\mathrm{s} \left[\left(1+\frac{R}{r_\mathrm{s}}\right)\left(1+\frac{R}{r_\mathrm{s}}\right)^2\right]^{-1},
\end{equation*}
with $r_\mathrm{s} = \unit[9.26^{+5.6}_{-4.2}]{kpc}$ from a fit to stellar data from the Milky Way \cite{Nesti:2013}.
Fig.~\ref{fig:DMProfiles} compares the shapes of these profiles. It is obvious that the current uncertainty on the true shape of the halo profile imposes large systematic uncertainties on any measurement of a DM annihilation cross section (or upper limits thereof) in the central Galactic DM halo -- not even taking into account uncertainties due to the complicated dynamics of baryonic matter in this region. Despite these large uncertainties, the central Galactic halo hosts the largest DM density of our Galaxy, which makes it a prime target for indirect DM detection.

\subsection{Searching for Dark Matter annihilations in the Galactic centre region}
In the last few years, several searches for DM signals were carried out using data of the Fermi-LAT instrument and ground-based $\gamma$-ray telescopes, leading to limits on the DM annihilation cross section $\langle\sigma v\rangle$ for DM masses in the range of about $\unit[10]{GeV}$ up to $\unit[10]{TeV}$. Good targets are dwarf galaxies \cite{Aharonian:2008,Albert:2008,Wood:2008,Aharonian:2009,Aliu:2009,Acciari:2010,Abdo:2010a,Aharonian:2011,Aleksic:2011,Aliu:2012,Aleksic:2014,Ackermann:2011,Ackermann:2014}, since they (i) consist mostly of DM and (ii) are therefore free of astrophysical $\gamma$-ray background, and galaxy clusters \cite{Aharonian:2009DM,Aharonian:2009Coma,Aleksic:2010Perseus,Acciari:20091275,Ackermann:2010Clusters,Abramowski:2012Fornax}. However, the astrophysical factors $J$ (cf.\ eq.~\ref{eq:CrossSection}) of these objects are generally small compared to what is obtained in observations of the central Galactic DM halo (but note that unresolved substructure in galaxy clusters may lead to a \emph{boost} of the annihilation rate by several orders of magnitude, see e.g.\ \cite{Pinzke:2011}). 

As opposed to dwarf galaxies or galaxy clusters, which are point-like objects for current $\gamma$-ray instruments, the inner Galactic DM halo is rather extended, calling for a proper optimisation of the search region (\emph{region of interest}, ROI) to maximise the discovery potential. As far as the putative DM signal is concerned, ROI optimisation should take into account both the (assumed) DM halo profile of the inner Galaxy and the available FoV of the instrument to maximise the value of $J\Delta\Omega$. However, the optimisation is significantly complicated due to the fact that $\gamma$-ray emission in the inner Galactic halo is dominated by astrophysical sources (such as the GC point source, diffuse emission along the GC ridge, or emission from the Fermi bubbles, see section \ref{sec:part1}). When searching for a DM signal, either sky regions populated by these sources have to be removed from the search, or their contribution to the total $\gamma$-ray flux has to be carefully modelled and subtracted to obtain the putative DM signal. Unfortunately, at the barycentre of the Galaxy, where the DM distribution is expected to be largest in conventional scenarios, the existence of the GC point source makes it virtually impossible to extract any underlying DM signal from the central $\sim\unit[10]{pc}$ region (if not the emission itself is considered to be of DM origin, see below). Compared to the $\unit{TeV}$ range, diffuse ridge emission is much more intense and rather extended at $\sim\unit{GeV}$ energies, such that this flux contribution is generally harder to take into account when searching for DM at these energies. Furthermore, large scale emission like that of the Fermi bubbles (if not itself of DM origin), must be accounted for in \unit{GeV} DM searches, whereas at \unit{TeV} energies such emission must not (yet) be considered due to the limited FoV of current instruments and the way hadronic background is subtracted.

However, even when avoiding the innermost region, $J$-factors for the inner ($\sim\unit[1]{kpc}$ in diameter) Galactic halo are of the order of $\unit[10^{22}-10^{25}]{GeV^2/cm^5}$,  typically a few orders of magnitude larger than what is found for dwarf galaxies or galaxy clusters. For the Galactic halo, the size of the ROI plays an important role, since the expected count rate in the detector scales with $J\Delta\Omega$, where $\Delta\Omega$ is the ROI size (see eq.~(\ref{eq:CrossSection})). As an example, for the NFW profile given in eq.~(\ref{eq:Profiles}), a H.E.S.S.\ search using a circular ROI of outer radius $1^\circ$ centred on \astar, where a region of Galactic latitudes $|b|\leq 0.3^\circ$ is removed, yields $J\sim\unit[3\times 10^{24}]{GeV^2/cm^5}$ and $J\Delta\Omega \sim \unit[2\times 10^{21}]{GeV^2 sr/cm^5}$ \cite{Abramowski:2011Halo, Abramowski:2013LineSearch}. In a similar study \cite{Gomez:2013}, using Fermi-LAT data with a large ROI radius of $16.7^\circ$, removing the region where $|b|\leq 0.6^\circ$, $J\sim \unit[10^{23}]{GeV^2/cm^5}$, but $J\Delta\Omega \sim \unit[3\times 10^{22}]{GeV^2 sr/cm^{5}}$ are found. A careful, \emph{a-priori} selection of the ROI is therefore important to obtain the most sensitive measurement for a given instrument.

\subsubsection{The Galactic Centre point source as a Dark Matter candidate}
Given that the DM density profile is expected to be strongly peaked towards the GC, one can ask the question whether or not the VHE $\gamma$-ray emission of the GC point source \hgc\ might be dominated by photons from the annihilation of TeV-mass DM particles. This question was addressed already early after the discovery of the emission (e.g.\ \cite{Hooper:2004,Aharonian:2006wh}). The fact that the emission is a point source for H.E.S.S.\  means that this scenario can only be realistically considered if a rather cuspy DM density profile is realised (otherwise, a more extended emission would be expected from standard halo profiles, cf.\ Fig.~\ref{fig:DMProfiles}). At the same time, the (average) $\gamma$-ray spectrum of such annihilations should match the observed energy distribution from the source. As discussed above, predicted energy spectra for $\gamma$-rays produced in DM annihilations into various standard model particles can be calculated rather precisely, and tend to be strongly curved (for reasons of energy conservation) as the energy approaches the the mass of the DM particle. Since the spectrum of \hgc\ is described by a powerlaw over most of its range, and only cuts off at energies of some $\unit[10]{TeV}$, typical annihilation spectra are disfavoured by the observation. In any case must the mass of the DM particle exceed \unit[10]{TeV}. 

The $\gamma$-ray emission from \hgc\ is therefore not easily explained as 
being dominantly produced by the most common DM
scenarios. As a consequence, the bulk of the 
$\gamma$-ray excess is probably of astrophysics rather than of particle
physics origin. However, a $\sim 10\%$ admixture of $\gamma$-rays from DM 
annihilations in the signal cannot be ruled out based on spectrum fits \cite{Aharonian:2006wh}.
Using this result and assuming an NFW-type halo profile, \cite{Aharonian:2006wh} calculated upper limits on the velocity-weighted annihilation cross section. These are at least two orders of magnitude above the thermal relic cross section, and thus do not put any constraints on the DM model parameter space.

\begin{figure}
\includegraphics[width=0.48\textwidth]{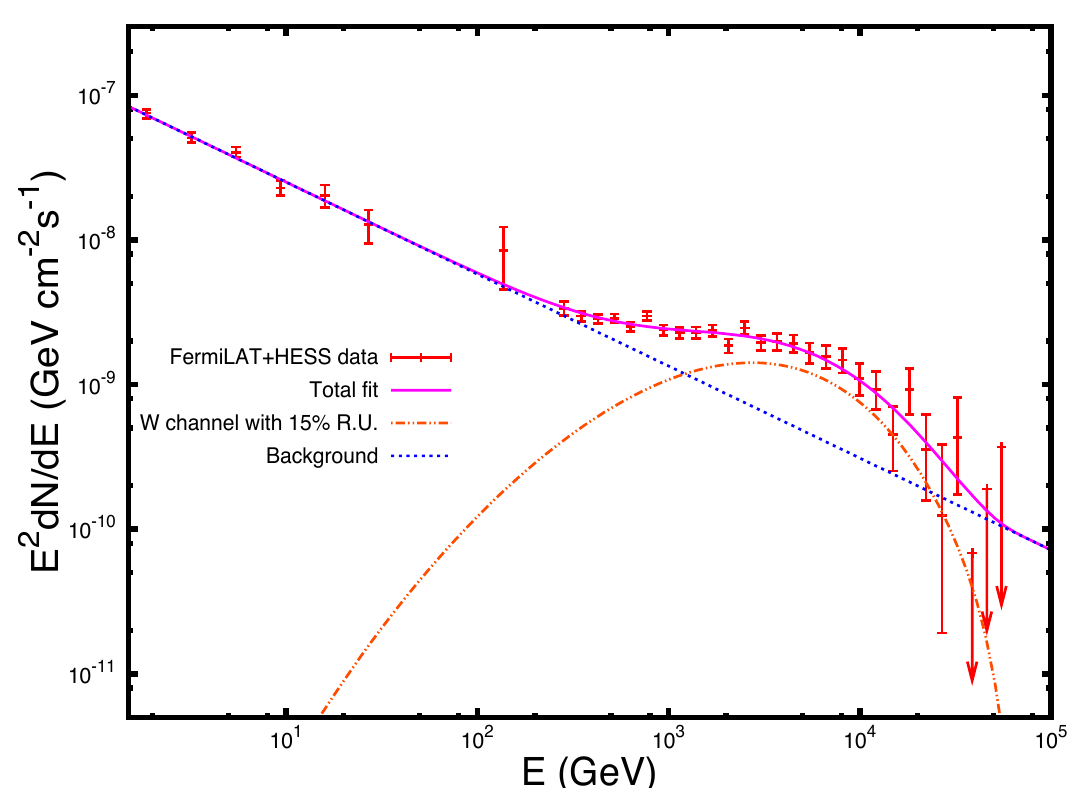}
\caption{Combined fit (full line) to the spectral energy distributions of the Fermi-LAT GC source 2FGL~J1745.6-2858 and the VHE source \hgc.  DM annihilation into a $W^+ W^-$ final state (dashed-dotted line) is shown on top of a powerlaw shaped background (dotted line). A best-fit DM mass of $\unit[51.7]{TeV}$ is found for this scenario. The background spectrum has a spectral index of $\Gamma=2.63$. Taken from \cite{Cembranos:2013}.}
\label{fig:GCPointSourceDMSpectrum}
\end{figure}

With spectral information on the Fermi-LAT GC source available (see \cite{Chernyakova:2011} and sect.~\ref{sec:HESource}), \cite{Cembranos:2013} show that the combined emission from 2FGL~J1745.6-2858 and \hgc\ can be described by DM annihilation into various final state particles on top of a powerlaw background (assumed to be of astrophysics origin). Fig.~\ref{fig:GCPointSourceDMSpectrum} shows as an example the fit of a $W^+ W^-$ final state spectrum, resulting in a DM mass of $\sim\unit[52]{GeV}$ and a underlying background spectrum with spectral index $\Gamma \sim 2.6$ (i.e.\ consistent with being of Galactic origin). It should be noted that for all final states a boost factor of the order $10^2-10^3$ over a standard NFW halo assumption is needed to explain the strong $\gamma$-ray signals (assuming that a thermal relic cross section, $\langle\sigma v\rangle \sim \unit[3\times 10^{-26}]{cm^3~s^{-1}}$ is realised). This means that the innermost part of the DM halo should be even more compressed than assumed in the NFW case, in agreement with the point-like morphology of the emission. No good fits to the $\gamma$-ray spectrum are found for leptonic final states.

\subsubsection{Dark Matter searches in the Galactic halo}
DM searches in the Galactic halo are the most promising way to either detect an annihilation signal or place rather strong limits on the annihilation cross section. As discussed above, this is because (i) the DM density, and hence the $J$ factor, is still large and (ii) astrophysical background by the GC point source is avoided. The solid angle $\Delta\Omega$ used for a halo search depends on the assumed halo profile and background by diffuse (astrophysical) emission along the Galactic Plane, but also on the available FOV of the instrument. Given typical diameters of IACT FOVs, DM halo searches at very high energies must therefore concentrate on the inner few degrees of the Galactic halo, whereas the Fermi-LAT instrument observes the entire sky and can therefore make use of a much larger part of the halo. On the other hand, at very high energies, astrophysical background due to diffuse emission is limited to Galactic latitudes smaller than $\sim 0.3^\circ$, whereas at GeV energies diffuse emission is observed to much larger scales, asking for an ROI more distant from the Galactic Plane.

\begin{figure}[htp]
\includegraphics[width=0.48\textwidth]{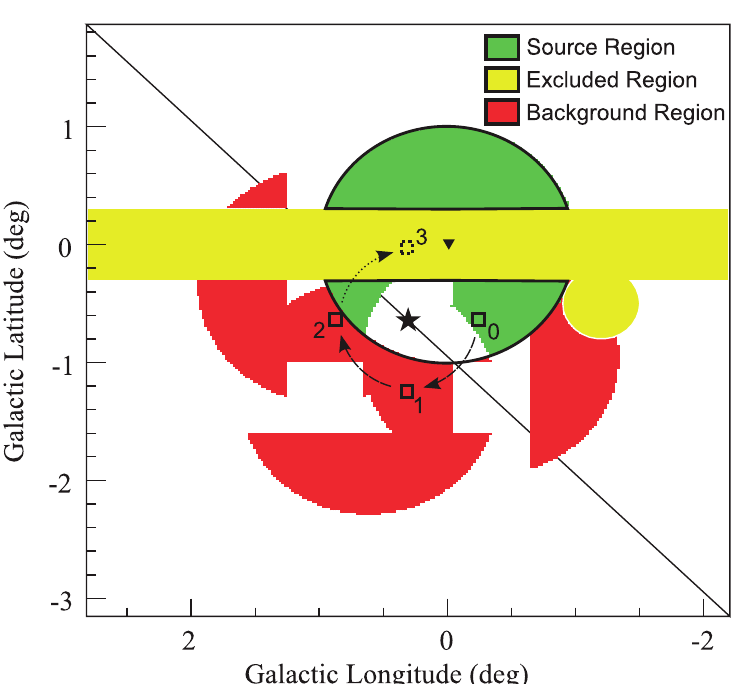}
\caption{Cosmic ray background subtraction technique used in the H.E.S.S.\ DM halo search \cite{Abramowski:2011Halo}. In this example, a telescope pointing position below the Galactic plane (depicted by the star) has been chosen. The ROI (the actual search region) is the green area inside the black contours, centred on the GC (black triangle). Yellow regions are excluded from the analysis because of contamination by astrophysical $\gamma$-ray sources. The cosmic ray background is estimated from the red areas, which are on average further away from the GC. This choice guarantees similar $\gamma$-ray detection efficiency in both the source and background regions, while the contribution to the $\gamma$-ray flux of putative DM annihilations is expected to be smaller in the background region due to the shape of the DM density profile.}
\label{Fig:SearchRegion}
\end{figure}

At very high energies, the so-far only generic search for annihilating DM in the Galactic halo was carried out by H.E.S.S. using $\sim\unit[100]{h}$ of GC observations \cite{Abramowski:2011Halo}. The search was carried out in an ROI of radius $1^\circ$ around the GC. To get rid of known astrophysical background, the Galactic plane was excluded from the search. Since ground-based observations with IACTs suffer from a large amount of background events by (isotropically distributed) charged cosmic rays (see sect.~\ref{sec:CurrentInstruments}), these have to be subtracted from the signal on a statistical basis. As the background varies strongly with the observation conditions, it is best evaluated using control regions from the same set of observations, i.e.\ sky regions where no $\gamma$-ray signal is expected, and hence only background events are recorded. For DM halo searches, however, this poses a problem, since the GC halo is much more extended than the FOV of a IACT, leaving no space to place control regions. The only possible way to deal with this situation is to take advantage of the steeply falling (with increasing galactocentric distance) DM density: the ROI is defined close to the GC (where the DM density is putatively large), whereas background control regions are placed further away from the GC (where the DM density is lower). The number of expected DM annihilation events is thus smaller in the background regions compared to the ROI. Still, when correcting for the background in the ROI, unavoidably part of the putative DM signal is subtracted, and only a residual annihilation signal is measured, somewhat limiting the sensitivity of the search. Fig.~\ref{Fig:SearchRegion} illustrates details of the applied procedure. 

\begin{figure}
\includegraphics[width=0.48\textwidth]{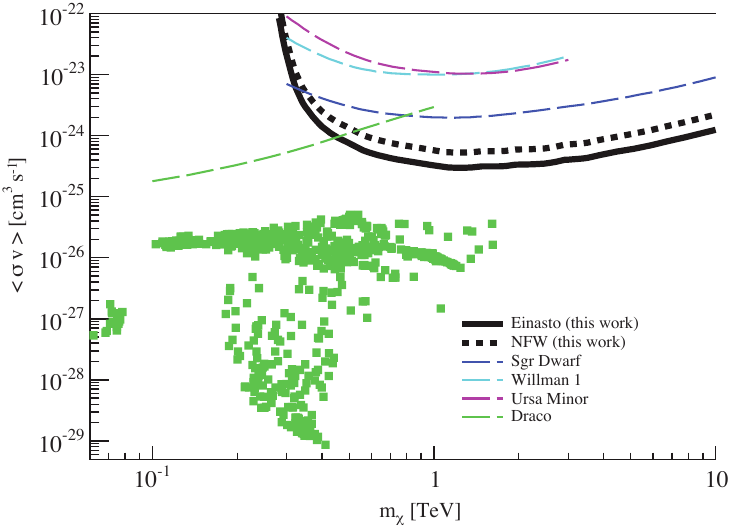}
\caption{Upper limits (95\% CL) on the DM annihilation cross section as a function of DM mass, extracted from H.E.S.S.\ observations of the inner Galactic halo. Generic annihilation into a quark-antiquark pair is assumed (using the annihilation spectrum given in \cite{Tasitsiomi:2002}), and two different DM halo profiles are investigated. Limits are compared to others derived from the observation of dwarf galaxies, and to cross section predictions calculated in the framework of mSUGRA. Taken from \cite{Abramowski:2011Halo}.}
\label{fig:HESSDMLimits}
\end{figure}

Since no $\gamma$-ray excess was observed in the ROI, upper limits on the annihilation cross section $\langle\sigma v\rangle$ were derived \cite{Abramowski:2011Halo}, assuming generic annihilation into quark-antiquark states. Fig.~\ref{fig:HESSDMLimits} shows these limits as a function of DM mass, compared to limits obtained from the observation of dwarf galaxies. The measurement provides the to date best limits for DM masses in the range $\sim\unit[0.5-10]{TeV}$. Still the limits are at least an order of magnitude away from cross section predictions of typical DM models.

DM searches with HE $\gamma$-rays have been conducted for a large variety of ROIs, and different results (both upper limits and tentative detections) have been reported. Shortly after the first LAT data became available, the detection of an extended (few degrees), roughly spherically-symmetric emission centred on the GC was reported \cite{Goodenough:2009,Hooper:2011}: after having accounted for diffuse emission from the Galactic disk and buldge, and for known point sources, a residual emission remains. The spectrum of the emission exhibits a peak at an energy of about $\unit[1-4]{GeV}$, which can be interpreted as a $\gamma$-ray signal from DM annihilating predominantly into $\tau$ lepton \cite{Hooper:2011} or $b\bar{b}$ \cite{Goodenough:2009} final states. The morphology of the excess is reasonably compatible with an NFW-type DM profile, and the annihilation cross section matches within a factor of a few that of a thermal relic. \cite{Hooper:2011a} improve on these early analyses and identify in three years of Fermi-LAT data a point-like $\gamma$-ray emission centred on the GC (with similar spectral and morphological properties as found by \cite{Chernyakova:2011}) which is embedded in an extended halo emission. As argued by the authors, another spectral component (besides the GC point source and diffuse emission) is needed to account for the total flux measured. Furthermore, they show that the residual emission from the extended region features a bump-like energy distribution peaking again at $\sim\unit[1]{GeV}$, which can be explained by DM annihilating either into lepton or hadron final states. The most recent analysis of this excess \cite{Daylan:2014} improves on these results both in terms of exposure time and angular resolution (by applying additional event selection cuts which improve the PSF). Astrophysical backgrounds are taken into account by means of template maps (including one accounting for the emission from the Fermi bubbles). The result of this study is that an introduction of a DM template significantly improves the fit of the data. The best fit results in a DM particle of mass $\sim\unit[31-40]{GeV}$ annihilating predominantly into $b\bar{b}$ pairs, and the best-fit annihilation cross section is about half the one required for a thermal relic, hence well consistent with cosmological predictions. 

Given the large density of astrophysical objects in the inner GC, it is clear that, in order to prove a DM origin of such a signal, all possible astrophysical  processes must be excluded as potential sources of the emission. As the emission extends to at least $\sim 10^\circ$ from the GC, \cite{Daylan:2014} argue that it is unlikely that the signal is produced by millisecond pulsars (see e.g.~\cite{Bednarek:2013}). However, as pointed out by \cite{Carlson:2014}, a population of cosmic ray protons injected into in the inner Galaxy $\sim\unit[1]{Myr}$ ago matches well the $\gamma$-ray excess attributed by \cite{Daylan:2014} to DM annihilation processes. In a similar study \cite{Petrovic:2014}, the emission is explained by inverse Compton emission from relativistic electrons injected into the GC at about \unit[1]{Myr} ago.

\begin{figure}
\includegraphics[width=0.48\textwidth]{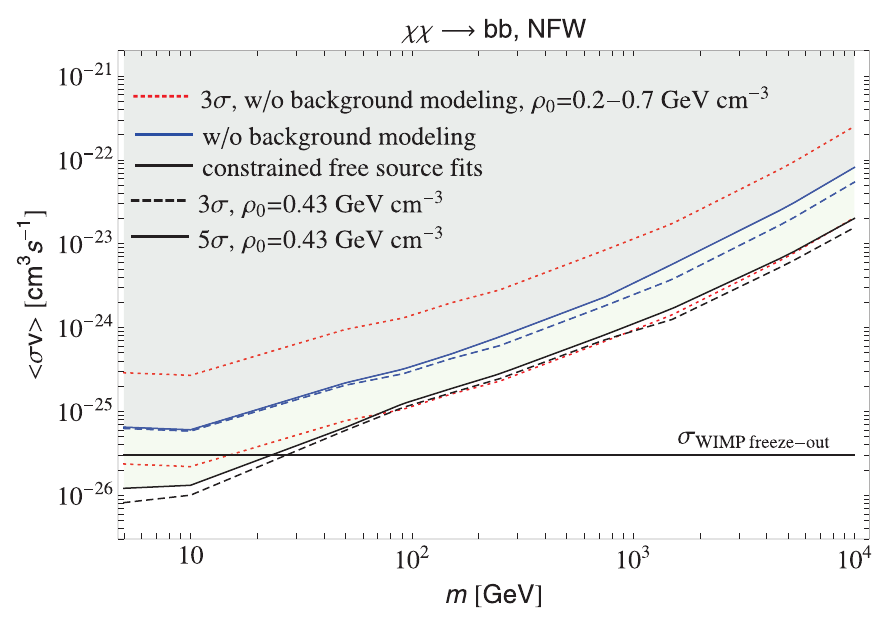}
\caption{Annihilation cross section limits from Fermi-LAT observations of the outer Galactic halo (at $\unit[3]{\sigma}$ and $\unit[5]{\sigma}$ significance, dashed and solid lines, respectively) for DM annihilation into $b\bar{b}$ pairs, assuming an NFW halo profile. Limits are given for the case where no diffuse emission modelling has been applied (blue curves, conservative limit) and with a diffuse $\gamma$-ray model in place (black curves). Figure taken from \cite{Ackermann:2012DM}.}
\label{fig:FermiLATDMLimits}
\end{figure}

The ROI chosen by the Fermi-LAT Collaboration to derive upper limits on DM annihilation in the Galactic halo \cite{Ackermann:2012DM} comprises large regions outside of the inner halo ($5^\circ < |b| < 15^\circ$, $|l| < 80^\circ$). This choice was made to avoid systematic uncertainties that come along with modelling the sources in the vicinity of the GC (as discussed above). Besides that, the main analysis is conservative in the sense that no astrophysical backgrounds are considered in the limit calculation (thus reducing potential systematics at the expense of worse, i.e.\ conservative, limits). Fig.~\ref{fig:FermiLATDMLimits} shows, as an example, the cross section limits (obtained from 24 month of Fermi-LAT data) for DM annihilation into $b\bar{b}$ pairs, assuming an NFW density profile. The limits cover a DM mass range from \unit[5]{GeV} up to \unit[10]{TeV}, and, at least for small DM masses, are close to the thermal relic cross section. The work of \cite{Gomez:2013}, instead, applies a H.E.S.S.-like ROI which is a concentric ring centred on the GC, excluding the Galactic Plane. Depending on the DM density profile considered, different inner and outer radii (the latter up to $\sim 17^\circ$) are chosen based on a signal-to-noise optimisation. Limits are found which are in general comparable to those found by \cite{Ackermann:2012DM}. However, \cite{Gomez:2013} additionally investigate the impact of a contracted NFW profile, for which much stronger limits are obtained.

\subsubsection{Searches for $\gamma$-ray lines and other spectral features}
\begin{figure}
\includegraphics[width=0.48\textwidth]{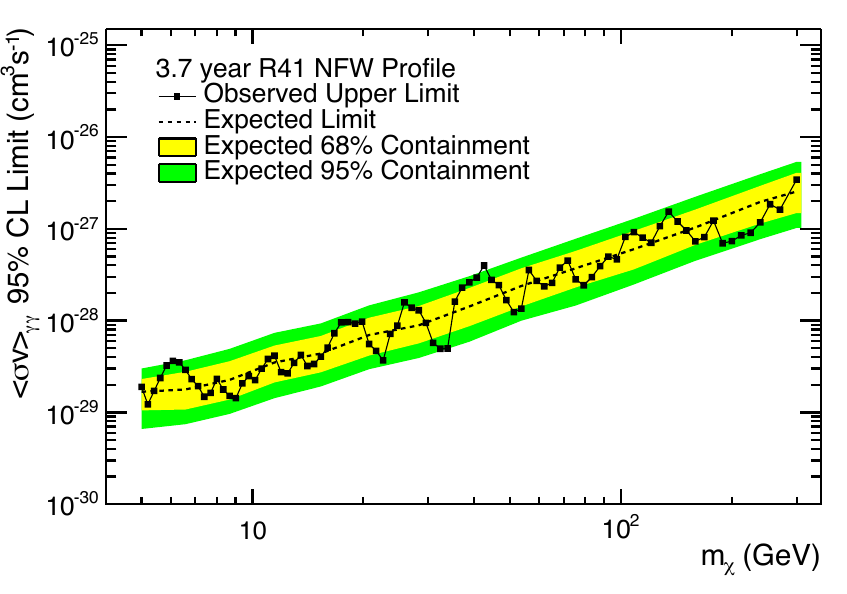}
\caption{Upper limits (95\% CL, 3.7 years of Fermi-LAT data) for DM annihilation into a two-photon final state as a function of DM mass. An NFW density profile is assumed. Upper limits are plotted as black markers. The shaded regions denote the 68\% and 95\% containment regions of the expected limit (obtained from simulations of the background statistics only). The dashed line shows the median expected limit for the background-only case. Figure taken from \cite{Ackermann:2013Lines}.}
\label{fig:FermiLineLimits}
\end{figure}

Several searches have been conducted with HE and VHE $\gamma$-rays to look for direct annihilation of the DM particles into a two-photon (and $\gamma Z$) final states. In a series of papers \cite{Abdo:2010Lines,Ackermann:2012Lines,Ackermann:2013Lines}, the Fermi-LAT collaboration presented cross section upper limits in a DM mass range of \unit[5--300]{GeV}. While \cite{Abdo:2010Lines} and \cite{Ackermann:2012Lines} use essentially the whole sky as an ROI (excluding only the Galactic Plane, but not the GC region), \cite{Ackermann:2013Lines} select circular ROIs centred on the GC, with the angular size of the ROI determined by signal-to-noise optimisation (which depends on the DM density profile considered). 
Latest annihilation cross section limits into the $\gamma\gamma$ final state are shown in Fig.~\ref{fig:FermiLineLimits} for an NFW profile. The limits reach $\unit[10^{-29}]{cm^3~s^{-1}}$, i.e.\ are more than three orders of magnitude lower than the thermal relic cross section. As the annihilation into two photons is loop-suppressed, and may therefore be suppressed by several orders of magnitude \cite{Bergstroem:1997}, these searches have just started to constrain the most favourite model predictions. A recent, complementary study addresses two-photon final states with $\gamma$-ray energies of $\unit[100]{MeV} - \unit[10]{GeV}$ \cite{Albert:2014}, and limits the mass range of gravitino DM.

The latest cross section limits put forward by the Fermi-LAT Collaboration are at tension with a recent detection \cite{Bringmann:2012b,Weniger:2012,Tempel:2012,Su:2012DM} of a putative $\gamma$-ray line feature at an energy of $\sim\unit[130]{GeV}$ from the inner Galaxy. Based on 43 month of Fermi-LAT data, \cite{Weniger:2012} identify this feature with a post-trial significance of $\unit[3.2]{\sigma}$; its annihilation cross section, assuming an NFW density profile, amounts to $\unit[2.27\times 10^{-27}]{cm^3~s^{-1}}$ ($\sim 10\%$ of the thermal relic cross section). This detection launched a large number of publications that interpret the feature in terms of a DM origin. However, it was soon found that the feature was not only present in the LAT Galactic halo data, but also seen in other datasets like e.g.\ gamma-ray events from the direction of the Earth's limb (see e.g.\ \cite{Hektor:2013}). Indeed did the Fermi-LAT collaboration recently show \cite{Ackermann:2013Lines} that the claimed signal is probably caused by a (not finally understood) instrumental effect, and that, using an improved energy reconstruction and taking statistical trials properly into account, the significance of the feature in the Galactic halo data drops to $1.6\sigma$.

\begin{figure}
\includegraphics[width=0.48\textwidth]{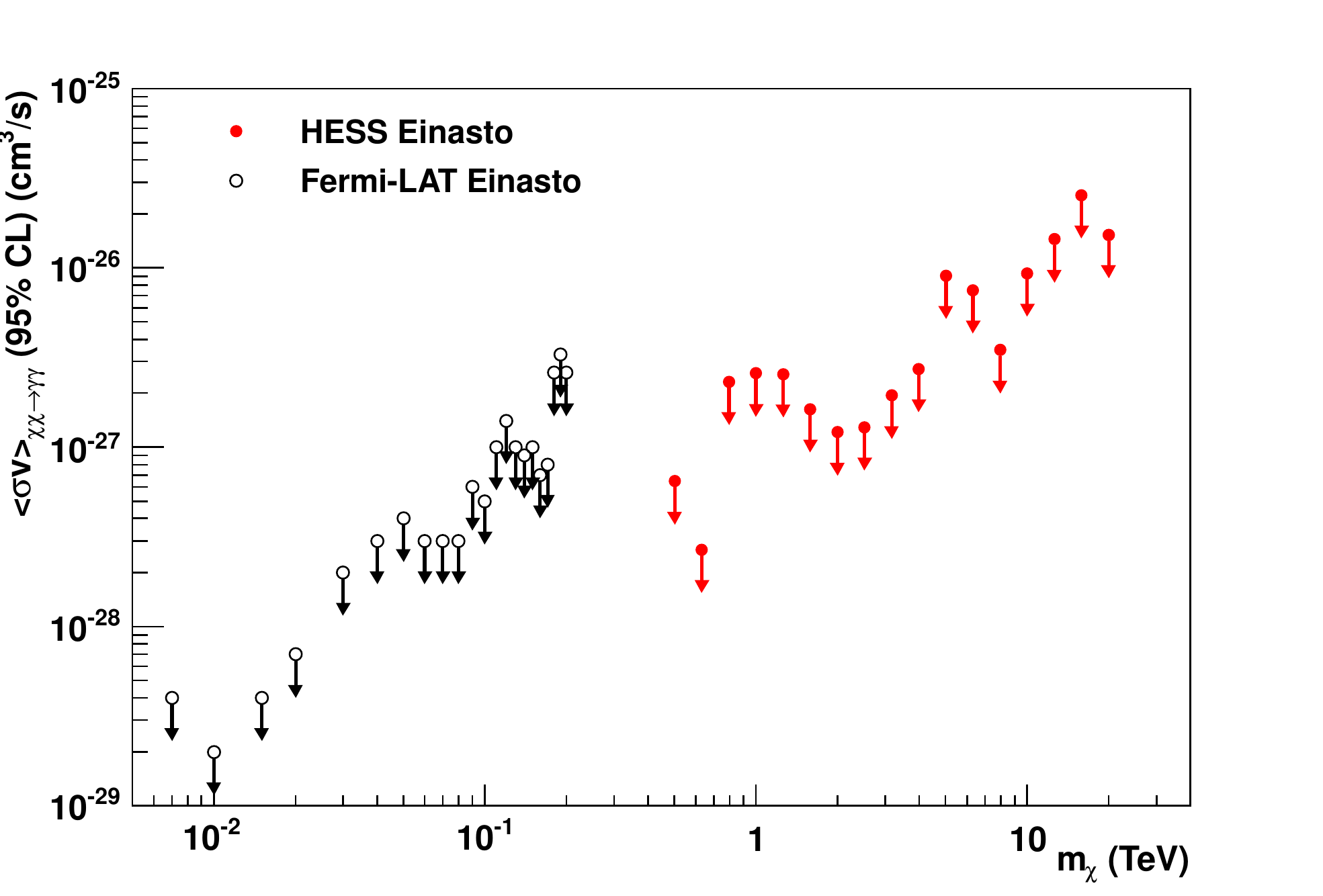}
\caption{Limits on the cross section for annihilation into two photons in the energy range $\unit[500]{GeV}-\unit[20]{TeV}$ as obtained by H.E.S.S. (red). Limits are assuming an Einasto density profile. Limits from the Fermi-LAT collaboration \cite{Abdo:2010Lines} are given for comparison. Figure taken from \cite{Abramowski:2013LineSearch}.}
\label{fig:FermiHESSLimits}
\end{figure}

Upper limits on two-photon states were also derived by the H.E.S.S.\ Collaboration for DM masses in the range of \unit[500]{GeV} to \unit[20]{TeV} \cite{Abramowski:2013LineSearch}. The search concentrated on a circular ROI of radius $1^\circ$ centred on the GC (similar to the one used for the continuum limits, cf.\ Fig.~\ref{Fig:SearchRegion}). The resulting limits are complementary to those derived with Fermi-LAT data and are presented in Fig.~\ref{fig:FermiHESSLimits}. The same type of technique can be used to search for internal bremsstrahlung features \cite{Bergstroem:2005a, Birkedal:2005,Bringmann:2008} in both HE and VHE data sets. Such searches have been carried out (e.g.\ \cite{Bringmann:2012b,Abramowski:2013LineSearch}), and corresponding limits have been placed.

\section{Conclusion and Prospects}
Ten years after the discovery of VHE $\gamma$-ray emission from the direction of the GC, observations of this region at energies of $\unit[100]{MeV}$ up to almost \unit[100]{TeV} with $\gamma$-ray satellites and ground-based Cherenkov telescopes became an important cornerstone in developing an improved understanding of the high-energy astrophysics of the inner region of the Milky Way. With the recent data from these instruments (in particular the Fermi-LAT and H.E.S.S.), a rich $\gamma$-ray morphology of the GC ridge becomes evident, providing strong indication for at least one, and presumably many more, particle accelerators within the inner few \unit[100]{pc} region, capable of accelerating particles to energies of at least some tens of TeV. Besides powering the CMZ, past or recent AGN activity of \astar\ or star forming episodes are probably responsible for driving a powerful outflow from the region that energises the giant Fermi bubbles recently detected above and below the Galactic Plane. At the same time, the GC region presumably hosts the largest concentration of dark matter particles in the Galaxy, and $\gamma$-ray observations are key to reveal the properties of the yet unknown DM particles. 

Despite this important progress, current $\gamma$-ray instruments still lack the sensitivity, angular and energy resolution to develop an even deeper understanding of the GC constituents and their interplay. Due to limited angular resolution, source confusion often prevents solid identification of $\gamma$-ray sources with counterparts at smaller wavelengths, and better sensitivity would enable more detailed investigations of source morphologies and temporal variability (besides detecting more sources). Especially at MeV to GeV energies, however, sensitivity is often limited by the presence (and still incomplete understanding) of diffuse $\gamma$-ray emission. It seems, however, clear that instruments with better resolution and sensitivity would be able to provide much more detailed information also on the diffuse emission, thereby further reducing systematic uncertainties in the modelling of this component.

Next generation $\gamma$-ray instruments like the GAMMA-400 space mission \cite{Galper:2013} and the ground-based Cherenkov Telescope Array (CTA, \cite{Acharya:2013}) will provide the necessary performance to mark a significant step forward in Galactic Centre astrophysics and dark matter searches, but also in high-energy astroparticle physics in general. Compared to existing instruments, both $\gamma$-ray telescopes will significantly improve in sensitivity, energy coverage, and energy as well as angular resolution. CTA, for which construction of a first partial array may start already in 2016, will cover an energy range from several $\unit[10]{GeV}$ to more than $\unit[100]{TeV}$. In its core energy range, it will outperform existing IACTs by a factor $\sim 10$ better flux sensitivity as well as improved angular resolution. The space-born GAMMA-400 instrument, currently planned for launch in 2019, will provide a smaller effective area than Fermi-LAT, but at the same time significantly improve in cosmic ray background rejection, and outperform Fermi-LAT by a large factor both in angular and energy resolution.

As construction of CTA is expected to start soon\footnote{CTA is designed to consist of two separate IACT arrays in the Northern and Southern Hermisphere, respectively. Here we restrict our discussion to the southern array, as it exhibits a superior view to the GC region.}, and the performance of the planned array is fairly well understood \cite{Bernloehr:2013}, we will take it as an example to sketch the impact of next generation instruments on the $\gamma$-ray view of the GC. 

Is the \gr\ point-like source \hgc\ connected to the supermassive BH \astar? As discussed in sect.~\ref{sec:SgrAEmission}, a smoking gun signal for a direct association would be the observation of correlated \gr/X-ray (or NIR) variability, with current constraints being rather weak (at the level of twice the quiescent \gr\ flux on time scales of $\sim 1\unit{hour}$). Assuming an increase in flux sensitivity of a factor of 10 compared to H.E.S.S., CTA would be able to rule out (or detect) such flares at the level of $\unit\sim 20\%$ of the quiescent state \gr\ flux. Another approach to identify the nature of \hgc\ is to search for
plausible candidate counterparts within the error circle of the
emission centroid (see sect.~\ref{sec:AEast}). Since the GC is a densely packed region, naively, both improved angular resolution and photon statistics help to disentangle different source contributions. For an expected angular resolution of  $\sim 0.03^\circ$ for energies in excess of \unit[1]{TeV} \cite{Bernloehr:2013}, and assuming that the statistical error on the emission centroid scales linearly with both the angular resolution and the flux sensitivity, a factor of $\sim 30$ improvement compared to H.E.S.S.\ is to be expected, leading to a statistical localisation uncertainty at the sub-arcsec level for a point-like \gr\ source. However, residual systematic uncertainties due to telescope deformations and atmospheric effects will make it  hard to push the total localisation uncertainty to values smaller than a few seconds of arc.

But is \hgc\ a point source after all? As shown by \cite{LindenProfumo:2012}, CTA will be able to distinguish between a point source emission scenario (e.g.\ \gr\ production close to \astar) and extended emission on angular scales of $\sim 0.02^\circ$ (e.g.\ caused by interactions of relativistic protons with the dense gas of the Circum Nuclear Disk). The authors show further that CTA angular resolution suffices to test whether the emission profile is radially symmetric or instead elongated in the direction of the Galactic Plane, which will help to disentangle astrophysical production processes from those assuming a DM origin. As CTA will extend the energy coverage down to a few $\unit[10]{GeV}$, and will therefore bridge a sensitivity gap to Fermi-LAT energies, it will be possible to precisely measure the spectrum of the GC point source across almost six orders of magnitude in energy, providing further constraints to emission models.

\begin{figure}
\begin{minipage}[b]{0.49\textwidth}
\includegraphics[width=\textwidth]{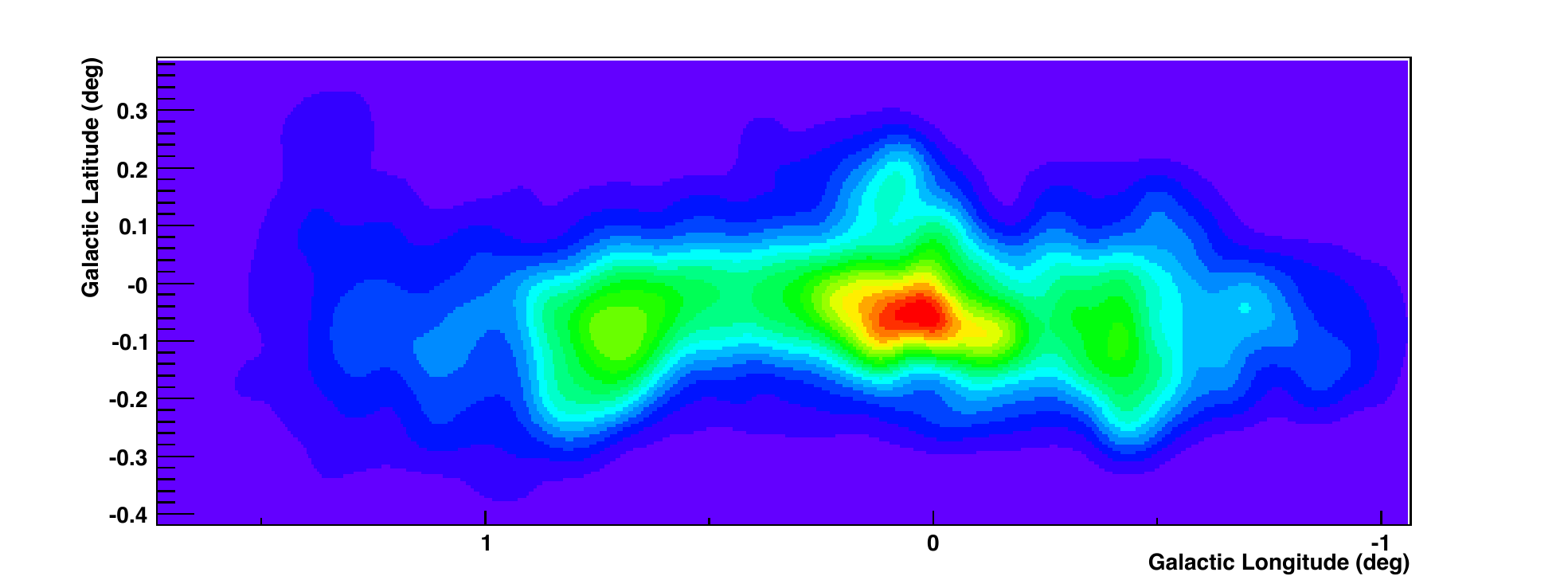}
\end{minipage}
\begin{minipage}[b]{0.49\textwidth}
\includegraphics[width=\textwidth]{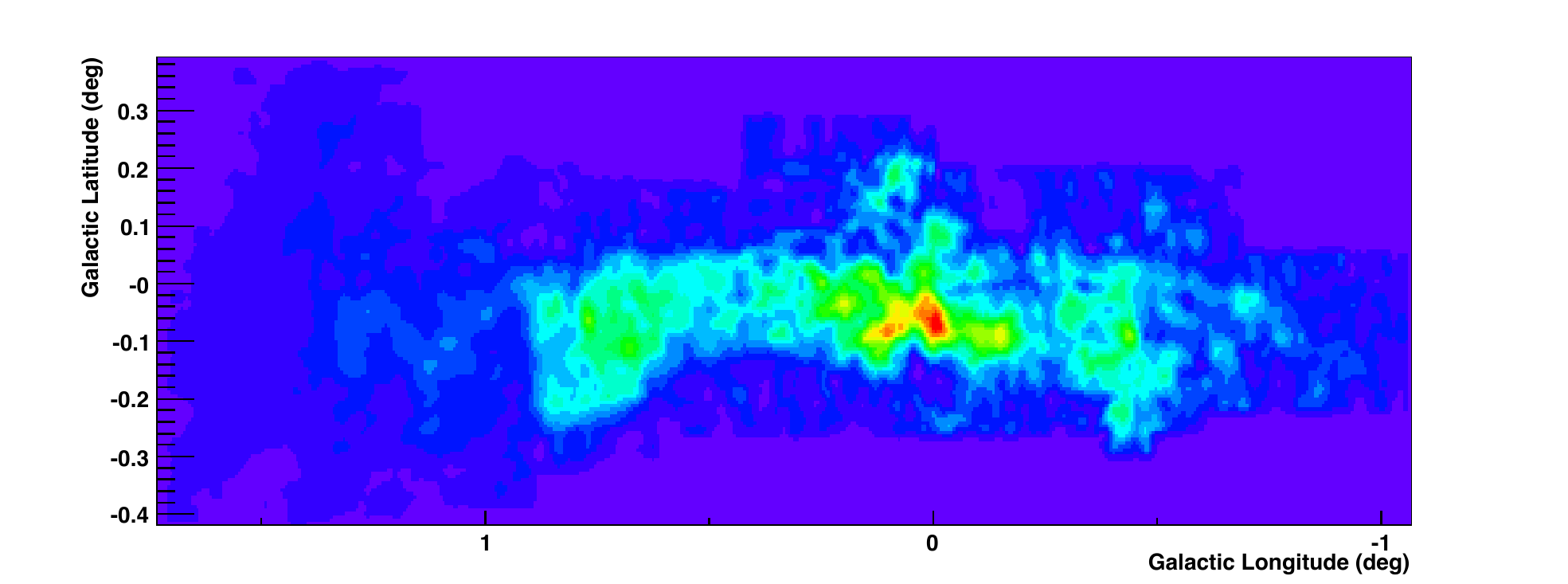}
\end{minipage} 
\label{fig:DiffuseCTA}
\caption{Comparison of simulated VHE \gr\ flux (in arbitrary units) of the GC diffuse emission as expected for H.E.S.S.\ (top, angular resolution $0.07^\circ$) and CTA (bottom, angular resolution $0.03^\circ$). The \gr\ emission is assumed to be produced by proton-gas interactions in the region's dense molecular clouds. Calculations assume that the protons got injected at the position of \astar\ $10^4$ years ago. Diffusion is governed by an (assumed energy-independent) diffusion coefficient $D=\unit[10^{30}]{cm^2 s^{-1}}$. Statistical fluctuations in the number of recorded \gr\ events are not taken into account.}
\end{figure}

CTA's superior angular resolution and sensitivity will also assist in understanding the properties of the diffuse emission, as it will allow to test the \gr-cloud correlation in much more detail than currently possible. Due to enhanced sensitivity, photon statistics will be sufficient to measure energy spectra for different parts of the CMZ. This will help to probe various scenarios of energy-dependent diffusion processes in the region, and also address the hypothesis of a possible existence of electron accelerators along the Galactic Centre ridge, which might or might not be responsible for the observed \gr\ emission in parts or in total. Fig.~\ref{fig:DiffuseCTA} shows a simple illustration of the improvement one might roughly expect with CTA in terms of mapping the morphology of the diffuse \gr\ emission. 

At the dark matter frontier, several studies have been carried out that show the expected performance of CTA to constrain DM models by Galactic DM halo observations. Given its much larger FoV of $\sim 8^\circ$ and its superior sensitivity as well as energy coverage, CTA can improve annihilation cross section limits placed by current IACTs by about one order of magnitude, and extend searches to a mass range of about $\unit[50]{GeV}-\unit[10]{TeV}$ (depending on the assumed particle spectrum, \cite{Doro:2013,Wood:2013,Pierre:2014,Silverwood:2014}). Similar to the analysis put forward by H.E.S.S. \cite{Abramowski:2011Halo}, these studies concentrate on the Galactic DM halo outside regions dominated by astrophysical \gr\ emission, and define control regions for the subtraction of cosmic ray background within the same field of view (but at larger distance to the GC than the signal region). Since the actual density profile within the DM halo is highly uncertain, predicted CTA limits on the annihilation cross section $\langle\sigma v\rangle$ can vary by several orders of magnitude (see, e.g., \cite{Pierre:2014}). Additionally, limits vary by a factor of a few depending on the assumed annihilation channel and the final array layout \cite{Doro:2013}, and recent studies show that an irreducible flux of Galactic diffuse electrons contributes a source of background that might reduce CTA's sensitivity by another factor of a few \cite{Silverwood:2014}.

\begin{figure}
\includegraphics[width=0.48\textwidth]{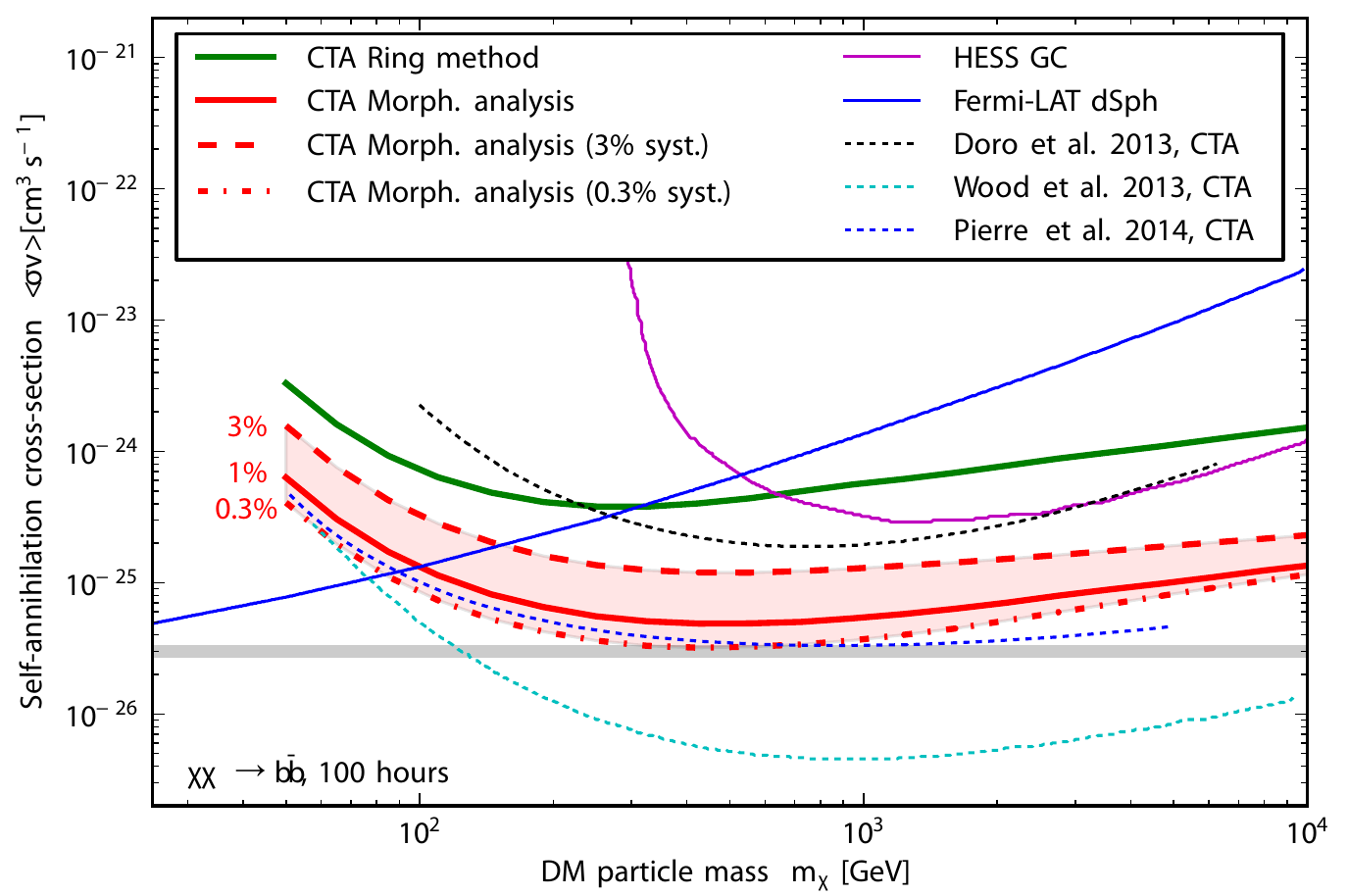}
\caption{Projected limits on the DM annihilation cross section $\langle\sigma v\rangle$ for CTA observations of the Galactic DM halo \cite{Doro:2013,Wood:2013,Pierre:2014,Silverwood:2014}. Thick lines represent the work of \cite{Silverwood:2014} from which the figure was taken. Existing limits from Fermi-LAT observations of dwarf spheroidal galaxies \cite{Ackermann:2014} and H.E.S.S.\ observations of the Galactic DM halo \cite{Abramowski:2011Halo} are shown for comparison. Limits are calculated for $\unit[100]{h}$ observation time, annihilation into $b\bar{b}$ final state, assuming an Einasto-type DM profile. Note that studies are based on different CTA array layouts. Only the work of \cite{Silverwood:2014} addresses the impact of the diffuse flux of Galactic electrons on the limits.}
\label{fig:CrossSectionComparison}
\end{figure}

Given these uncertainties, it is quite challanging to predict in detail CTA's potential to exclude concrete realisations of DM in particle physics models. Rather, the projected cross section limits are usually compared to the thermal cross section $\langle\sigma v\rangle=\unit[3\cdot 10^{26}]{cm^{3}s^{-1}}$. Fig.~\ref{fig:CrossSectionComparison} shows a compilation of selected CTA limit predictions, all assuming an observation time of $\unit[100]{h}$, an Einasto-like DM density profile and annihilation of the DM particles into $b\bar{b}$ pairs. If systematics are well under control, CTA will be able to significantly extend searches to smaller DM masses and improve existing limits by one order of magnitude for DM masses larger than a few $\unit[100]{GeV}$. CTA will thus be able to exclude DM models which predict annihilation cross sections of the order of the thermal cross section for a broad range of DM particle masses. For so-called contracted DM profiles and/or DM annihilation into tau leptons, predicted limits are even smaller by one to two orders of magnitude. However, it is clear that a solid contribution of CTA to DM annihilation searches in the Galactic halo requires both a very large observation time and a careful understanding of irreducible backgrounds and residual systematics.

\section*{Acknowledgement}
The author wants to thank the anonymous referee for valuable comments and fruitful discussions during the review process.

\bibliography{mybibfile}

\end{document}